\documentclass[]{mn2e}
\newcommand{\nab}{\mbox{\boldmath $\nabla$}}

\newcommand{\varphat}{\hat{\mbox{\boldmath $\varphi$}}}

\newcommand\bb[1]{\bf{#1}}

\newcommand{\bcdot}{\bb{\cdot}}

\usepackage{graphicx}
\usepackage{amsmath}
\usepackage{caption}
\usepackage{float}
\usepackage{color}
\graphicspath{{figures/}}

\begin{document}

\title[On the energy dissipation rate at the inner edge of
circumbinary discs]{On the energy dissipation rate at the inner edge
  of circumbinary discs}

\author[C. Terquem, J. C. B. Papaloizou] { Caroline
  Terquem$^{1,2}$\thanks{E-mail: caroline.terquem@physics.ox.ac.uk,
    J.C.B.Papaloizou@damtp.cam.ac.uk} and John C.
  B. Papaloizou$^{3}$\footnotemark[1] 
  \\
  $^{1}$ Physics Department,
  University of Oxford, Keble Road, Oxford OX1 3RH, UK \\
  $^{2}$ Institut d'Astrophysique de Paris, UPMC Univ Paris 06, CNRS,
  UMR7095, 98 bis bd Arago, F-75014, Paris, France \\
  $^{3}$ DAMTP, University of Cambridge, Wilberforce Road, Cambridge
  CB3 0WA, UK}

\date{}

\pagerange{\pageref{firstpage}--\pageref{lastpage}} \pubyear{}

\maketitle

\label{firstpage}

%
%===========================================================
%

\begin{abstract}
% 250 words max
  We study, by means of numerical simulations and analysis, the
  details of the accretion process from a disc onto a binary system.
  We show that energy is dissipated at the edge of a circumbinary disc
  and this is associated with the tidal torque that maintains the
  cavity: angular momentum is transferred from the binary to the disc
  through the action of compressional shocks and viscous friction.
  These shocks can be viewed as being produced by fluid elements which
  drift into the cavity and, before being accreted, are accelerated
  onto trajectories that send them back to impact the disc.  The rate
  of energy dissipation is approximately equal to the product of
  potential energy per unit mass at the disc's inner edge and the
  accretion rate, estimated from the disc parameters just beyond the
  cavity edge, that would occur without the binary. For very thin
  discs, the actual accretion rate onto the binary may be
  significantly less.  We calculate the energy emitted by a
  circumbinary disc taking into account energy dissipation at the
  inner edge and also irradiation arising there from reprocessing of
  light from the stars.  We find that, for tight PMS binaries, the SED
  is dominated by emission from the inner edge at wavelengths between
  1--4 and 10~$\mu$m.  This may apply to systems like CoRoT~223992193
  and V1481~Ori.

\end{abstract}


%
%===========================================================
%

\section{Introduction}
\label{sec:intro}

About 50\% of pre--main sequence (PMS) stars in T~associations have
been found to be in binary systems (Duch\^ene \& Kraus 2013).
Therefore, a large number of PMS stars evolve through accretion from
circumbinary discs, and those discs are also the birthplace of
planets.  

Discs around PMS binary systems were first detected at the same time
as discs around single stars in the late 1980s and 1990s through the
excess emission in the infrared and at submillimetric wavelengths
(Bertout, Basri \& Bouvier~1988). The first circumbinary disc to be
imaged, using interferometry and then adaptive optics, was that around
GG~Tau (Dutrey, Guilloteau \& Simon~1994, Roddier et al.~1996).  In
1997, Jensen \& Mathieu reported a deficit of near--infrared emission
from a few short--period PMS binaries, confirming the theoretical
prediction that binaries clear up a cavity in their circumbinary disc
(Lin \& Papaloizou 1979).  Recently, several circumbinary discs have
been imaged with ALMA (Czekala et al. 2015, 2016), and it is believed
that ALMA may be able to detect structures in circumbinary discs
produced by the tidal potential of the binary (Ruge et a. 2015).

Numerical simulations have shown that the cavity cleared up by the
binary does not prevent accretion onto the stars (Artymowicz \&
Lubow~1996), as gas spirals in along streams. Those streams connect to
circumstellar discs that ultimately are accreted onto the stars.
Such circumstellar discs in binary systems have been imaged using
adaptive optics (Mayama et al. 2010) and gas in the cavity of the disc
around GG~Tau has been imaged with ALMA (Dutrey et al. 2014).
 
Because of the presence of streams and circumstellar discs in the
cavity being fed by the circumbinary disc, PMS binaries are complex
systems.  Emission from such systems is indeed observed to have
contribution from various sources and to often be variable, but it is
not always understood what produces the emission and the variability
(Messina et al. 2016, Ardila et al. 2015, Gillen et al. 2014, Bary \&
Petersen~2014, Boden et al. 2009, Jensen et al. 2007).  Here we
investigate a new source of (variable) emission, namely energy
emission from the inner edge of the cavity.

Although the clearing of a cavity by the binary has been extensively
studied, the physical conditions in the region of the cavity edge have
yet to be elucidated.  Here we show that the action of the
gravitational torque from the binary, which truncates the disc and
produces the cavity, is accompanied by energy dissipation at the
disc's inner edge.  This energy dissipation, for the most part, comes
from shocks produced by fluid elements which are ultimately accreted
by the binary, and that suffer some oscillations in their distance to
the centre of mass of the binary resulting in them impacting the
disc's inner edge on their way in.  In addition, the edge of the
cavity may behave as a hard wall which absorbs the radiation from the
stars and re--emits it at longer wavelength.  These two sources of
energy produce an excess of emission which is in the mid--infrared for
spectroscopic binaries.

The plan of the paper is as follows. In section~\ref{sec:hydro_sim},
we present hydrodynamic simulations of a disc orbiting around a binary
system.  We calculate the rate of energy dissipation associated with
shocks which are found to be highly localised at the inner edge of the
disc.  As already noted by previous authors, we find that the rate of
accretion onto the binary {tends to be suppressed, compared to what
  would be expected from the value of the surface density just beyond
  the cavity edge in the absence of the binary, } when the disc's
aspect ratio decreases below 0.05 or so.  However, the energy
dissipation at the disc's inner edge is not affected by this
suppression.  In section~\ref{sec:analysis}, we develop an analysis to
elucidate the origin of this energy dissipation.  We show that it is
associated with the tidal torque which maintains the cavity: angular
momentum is transferred from the binary to the disc through the action
of compressional shocks and viscous friction.  The rate of energy
dissipation is given by the binary angular velocity times the torque
exerted by the binary, and does not directly depend on the actual
accretion rate onto the binary. In section~\ref{sec:particle_sim}, we
present particle simulations to illustrate the dynamics near the
disc's inner edge in more detail and the accretion flow onto the
binary.  These simulations show that, before being accreted, particles
are in general accelerated onto a trajectory that sends them back
towards the circumbinary disc. Shocks that result at the disc's inner
edge together with the action of viscous friction circularize the
orbits of the particles and dissipate energy.  We find that the energy
released during one collision of a unit mass particle with the disc's
inner edge is on the order of the potential energy per unit mass
there, leading to agreement with the results of the hydrodynamic
simulations.  When accretion onto the binary is reduced, the particles
which end up being accreted undergo more collisions, such that the
resulting total energy dissipation rate stays the same.  In
section~\ref{sec:SED}, we calculate the spectral energy distribution
(SED) of a circumbinary disc taking into account the energy dissipated
at the inner edge and also irradiation of the inner edge by the stars.
We show that, for tight binaries having a separation
$\sim 10$~R$_{\sun}$, emission from the edge completely dominates the
SED in the mid--infrared (from 1--4 to 10~$\mu$m).  In
section~\ref{sec:discussion}, we summarize and discuss our results.
We conclude that the processes presented in this paper may help to
explain the excess of emission in the mid--infrared observed in some
PMS binary systems, like CoRoT~223992193 (Gillen et al. 2014) and
V1481~Ori (Messina et al. 2016).

%
%===========================================================
%

\section{Hydrodynamic simulations}
\label{sec:hydro_sim}

We have performed two dimensional hydrodynamic simulations of 
a disc orbiting exterior to a binary system in circular orbit using NIRVANA
(see, e.g., Ziegler \& Yorke 1997).  This code has been frequently used to simulate
discs interaction with orbiting bodies (e.g., Nelson et al. 2000).

\subsection{Governing equations }

The governing equations solved numerically express the conservation of
mass and momentum for a razor thin disc.  In a non rotating frame with
origin at the location of the centre of mass of the binary, these take
the form:

\begin{equation}
  \frac{\partial \Sigma}{\partial t} + \nab {\bcdot} (\Sigma \bb{v})
  = 0 \, \label{contg} , 
\end{equation}
\begin{equation}
  \Sigma \left[ \frac{\partial \bb{v}}{\partial t} +  \left( \bb{v}
  {\bcdot} \nab \right) \bb{v} \right]  =    -\Sigma \nab \Phi   +
                                           \nab {\bcdot}  {\bf T}
%                                           \textcolor{blue}
                                           { +
                                           \Sigma {\bf F}} \, \label{mog} ,
% \frac{\partial { E} }{\partial t} + \del {\bcdot} (E \bb{v}) &=&
% -{\Pi} \del {\bcdot }({\bf v}) + \epsilon_v - 2 F_{+} - \nabla \cdot
% (2H F{\hat R})+ S_{E}, \label{engg} ,
% \frac{\partial P}{\partial t} + \bb{v} \bcdot \del P &=& -\gamma
% P\del \bcdot \bb{v}\,\label{adeq}
% \frac{\partial \bb{B}}{\partial t} &=& \del \btimes ( \bb{v} \btimes
% \bb{B} - \eta \del \btimes \bb{B} ) \, \label{induct} .
%\label{basic_eq}
\end{equation}

\noindent where $\Sigma$ is the surface density, ${\bf v}$ is the
velocity, ${\bf T}$ is the stress tensor which has contributions from
the vertically integrated pressure, $\Pi,$ and the standard
Navier--Stokes viscous stress tensor. The gravitational potential is
$\Phi$, and a drag force per unit mass {\bf F} is also included. Note
that the latter is set to zero in our grid based simulations but may
be retained in our analytic discussion given in section~\ref{Tidanal}
below.  In the simulations presented below, we will use
an~$\alpha$--prescription for the kinematic viscosity (Shakura \&
Sunyaev~1973).

\subsection{Computational setup and initial conditions}\label{setup}

We adopt a system of dimensionless units for which the total mass of
the binary system $M_1+M_2$ is unity.  In these units, the mass of the
primary is $M_1= 7/12$ and the mass of the secondary is $M_2 = 5/12.$
We choose the unit of distance such that the separation, $D,$ of the
binary components, taken to be in circular orbit, is unity and the
unit of time to be such that the binary angular frequency
$\omega = \sqrt{G(M_1+M_2)/D^3}$ is unity.  Thus, in these units, the
binary orbital period is $2\pi.$ In a system of dimensionless
cylindrical coordinates $(r, \varphi),$ the radial computational
domain is taken to be $[ r_{\rm i},r_{\rm o}]$ with $r_{\rm i} =0.7 $
and $r_{\rm o} = 10.5.$ Thus the binary orbits interior to the
domain. The $\varphi$ domain is $[0, 2\pi].$ The boundary condition is
taken to be inflow at $r=r_{\rm i}$ with the boundary at $r=r_{\rm o}$
taken to be rigid.  We have performed simulations on an equally spaced
grid with $N_r$ grid points in the radial direction and $N_{\varphi}$
grid points in the azimuthal direction.  For the simulations presented
here we have adopted $N_r =384$ and $N_{\varphi} = 512$ and checked
that the same behaviour is obtained adopting twice the resolution in
both $r$ and $\varphi.$

The discs are assumed to be locally isothermal with a constant aspect
ratio $H/r.$ Thus $\Pi = \Sigma H^2\Omega^2,$ where $\Omega$ is the
angular velocity.  For kinematic viscosity, $\nu,$ we adopt an
$\alpha$ prescription (Shakura \& Sunyaev 1973) with $\alpha$ taken to
be constant, i.e.  $\nu = \alpha H^2 \Omega$. 
\begin{table}
%\begin{centering}
\begin{tabular}{ccc}
\hline
 Model & $H/r$ & $\alpha$ \\
\hline
 A  & $0.1$ & $0.01$\\
 B  & $0.07$ & $0.02$\\
 C  & $0.05$ & $0.04$ \\
 D  & $0.05$ & $0.02$ \\
\hline
\end{tabular}
\caption{Parameters of models for  which results are described in the text.
The first column gives the model label, the second the adopted disc aspect ratio
and the third gives the value of $\alpha.$
Models A, B and C  have the same effective kinematic viscosity while
for model D this is a factor of two smaller
\label{table1}}
\end{table}
The
value of these parameters for the models considered here are given in
table \ref{table1}.  In order to handle shocks an artificial viscosity
was adopted (see e.g. Stone \& Norman 1992 ).

For initial surface density profile we adopt
$\Sigma = \Sigma_0 (r/2.5)^2$ for $r < 2.5$ and
$\Sigma = \Sigma_0(2.5/r)$ for $ r> 2.5,$ where $\Sigma_0$ is a
constant.  This is taken to be $9 \times 10^{-4}$ in dimensionless
units.  This form exhibits a cavity, of the type eventually maintained
in the simulations, with radius $r \sim 2.5.$ Note that, as self--gravity
is neglected and the binary is taken to be in a fixed circular orbit,
this value of $\Sigma_0$ is arbitrary.  Thus an arbitrary constant
scaling may be applied to the surface density, disc mass and
quantities such as the rate of energy dissipation in the disc
presented below.

\subsection{Numerical results}

\begin{figure}
\begin{center}
%\vspace{-0cm}
%\hspace{-2cm}
\includegraphics[scale=0.31]{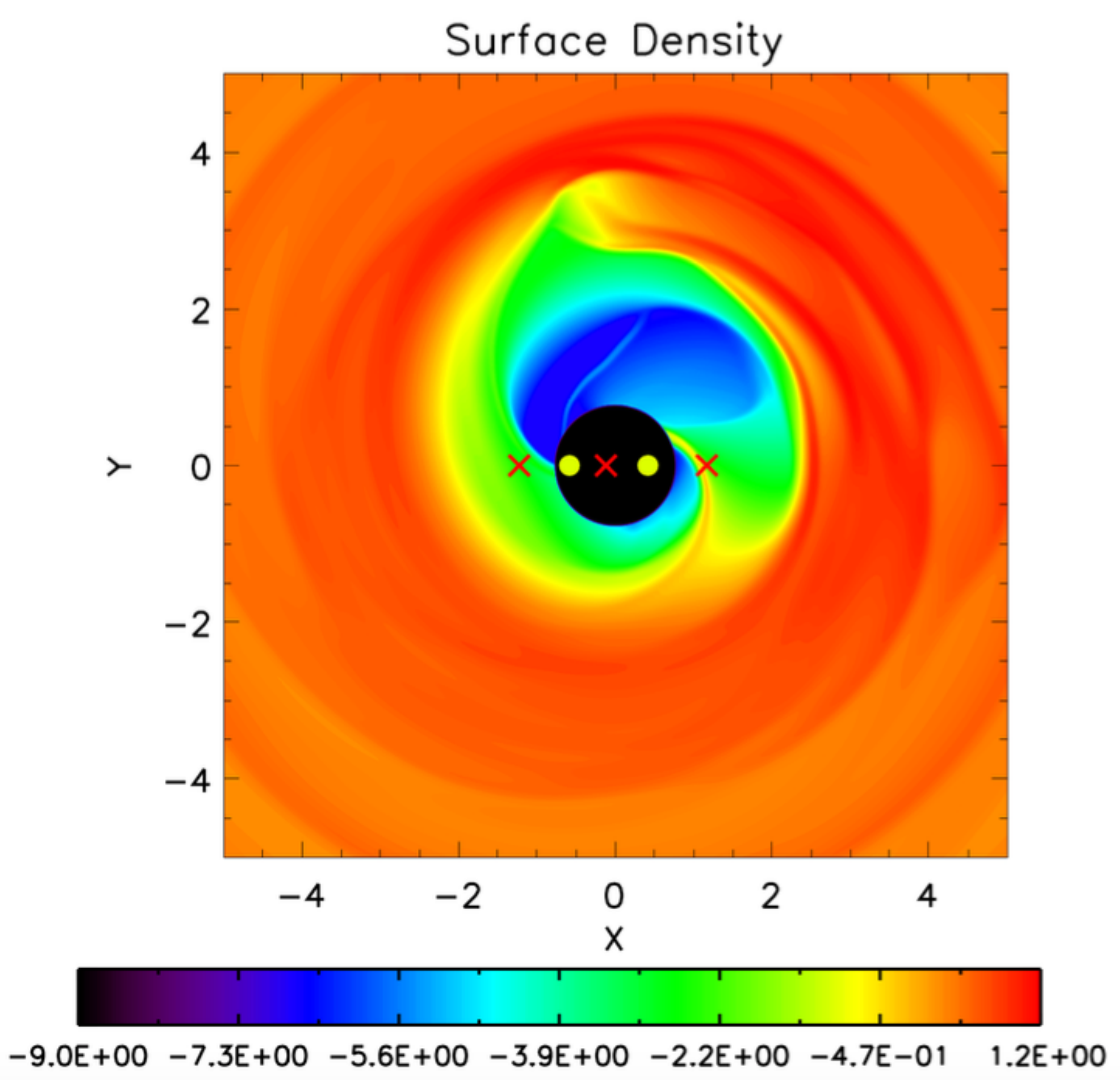}
%\vspace{3cm}\hspace{-0cm}
\includegraphics[scale=0.31]{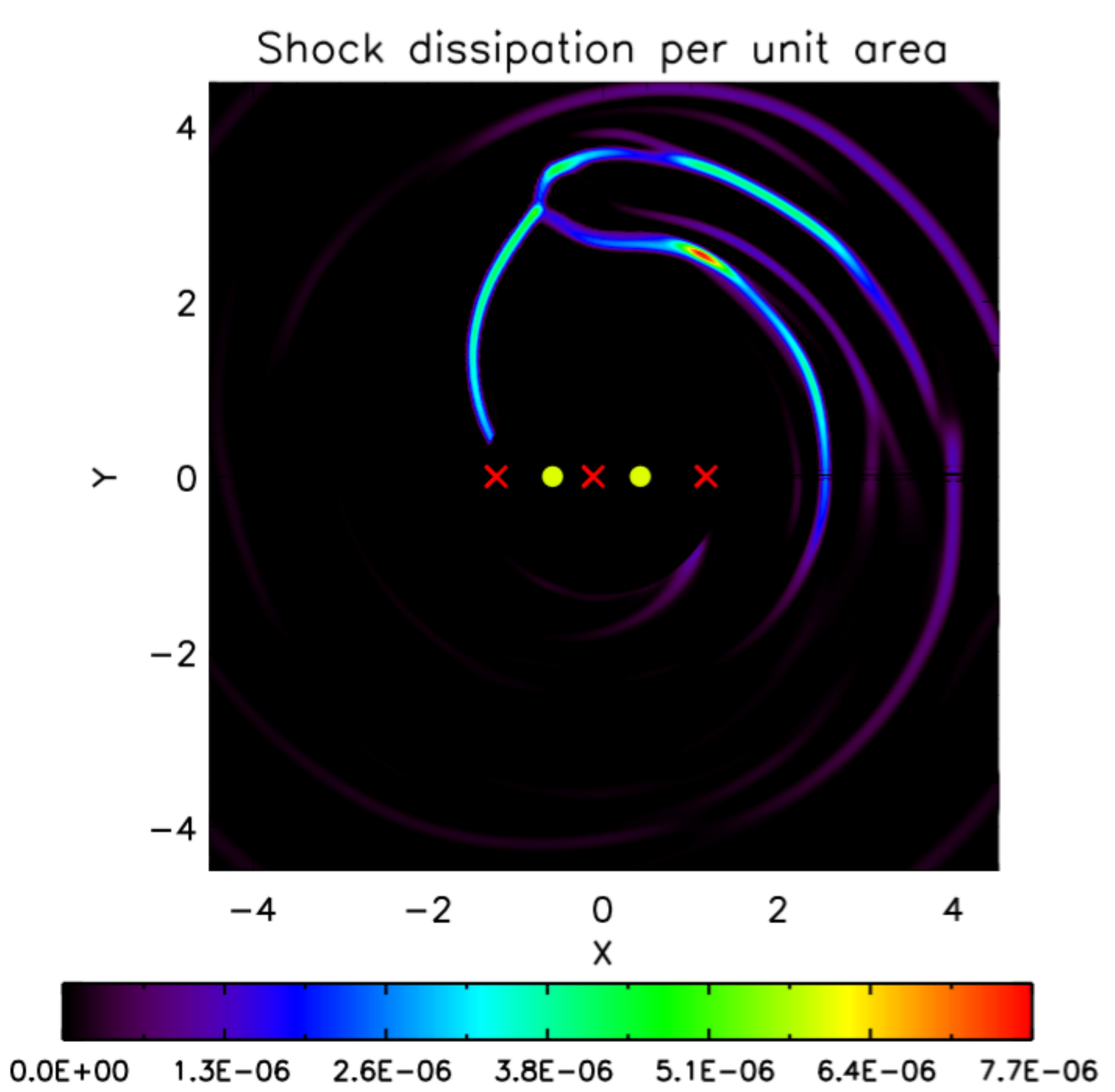}
\caption{The upper panel shows a contour plot of
  $\log_{10}(10^4\Sigma),$ with $\Sigma$ expressed in dimensionless
  units, for model A, after $208$ binary orbits.  The lower panel
  represents the rate of dissipation of energy we associate with
  shocks (see text).  In both panels, the yellow circles near the
  centre indicate the centres of the stellar components and the red
  crosses indicate the three collinear Lagrangian points.
  %\textcolor{blue}
  {The primary star is on the right with coordinates
    $(0.41,0)$, whereas the secondary star is on the left with
    coordinates $(-0.58,0)$.  The Lagrangian points are $L_3$, $L_1$
    and $L_2$ from right to left.  }}
\label{INCHHRVISC208}
\end{center}
\end{figure}

\begin{figure}
\begin{center}
%\vspace{-0cm}
%\hspace{-2cm}
\includegraphics[scale=0.31]{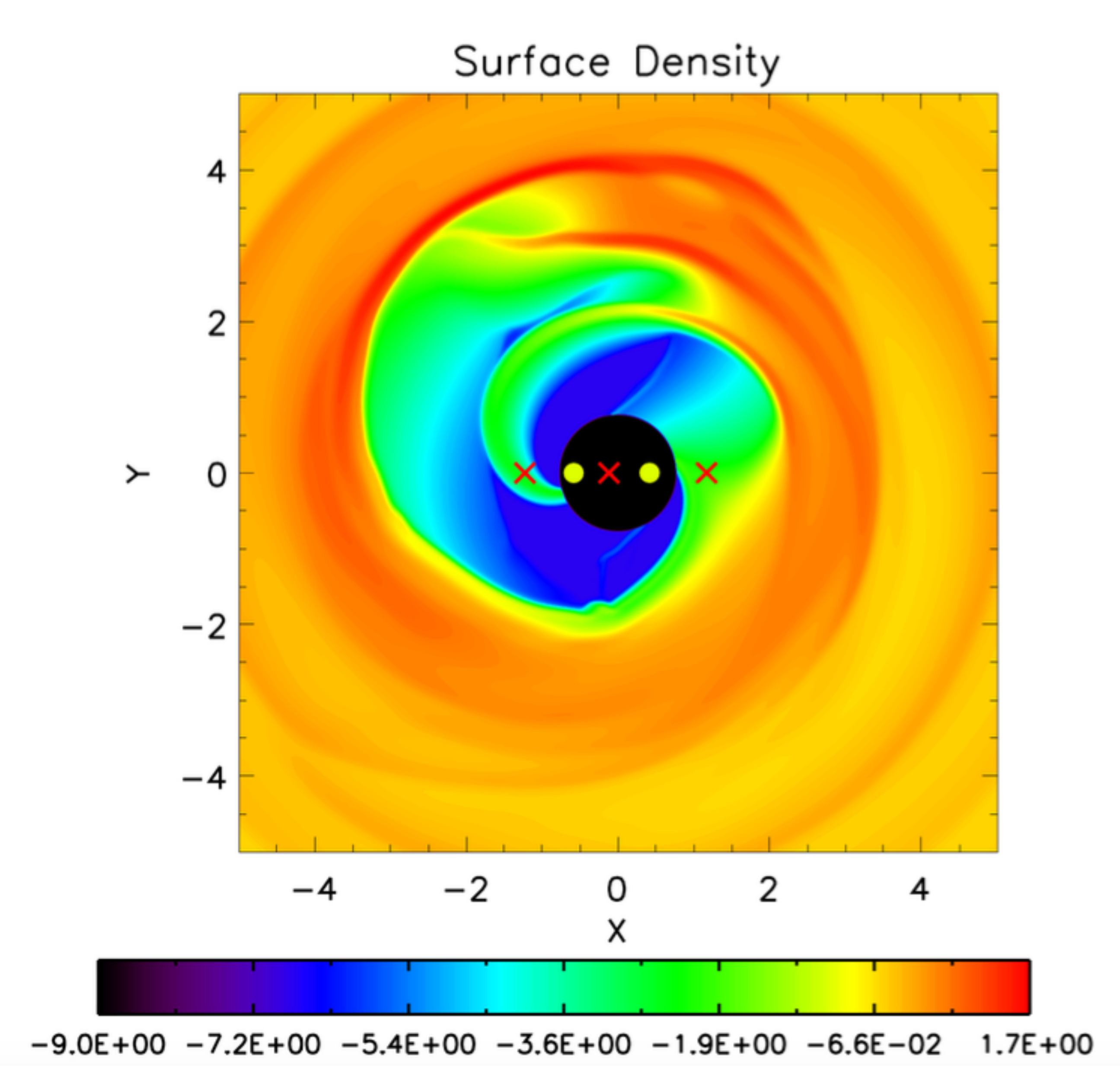}
%\vspace{3cm}\hspace{-0cm}
\includegraphics[scale=0.31]{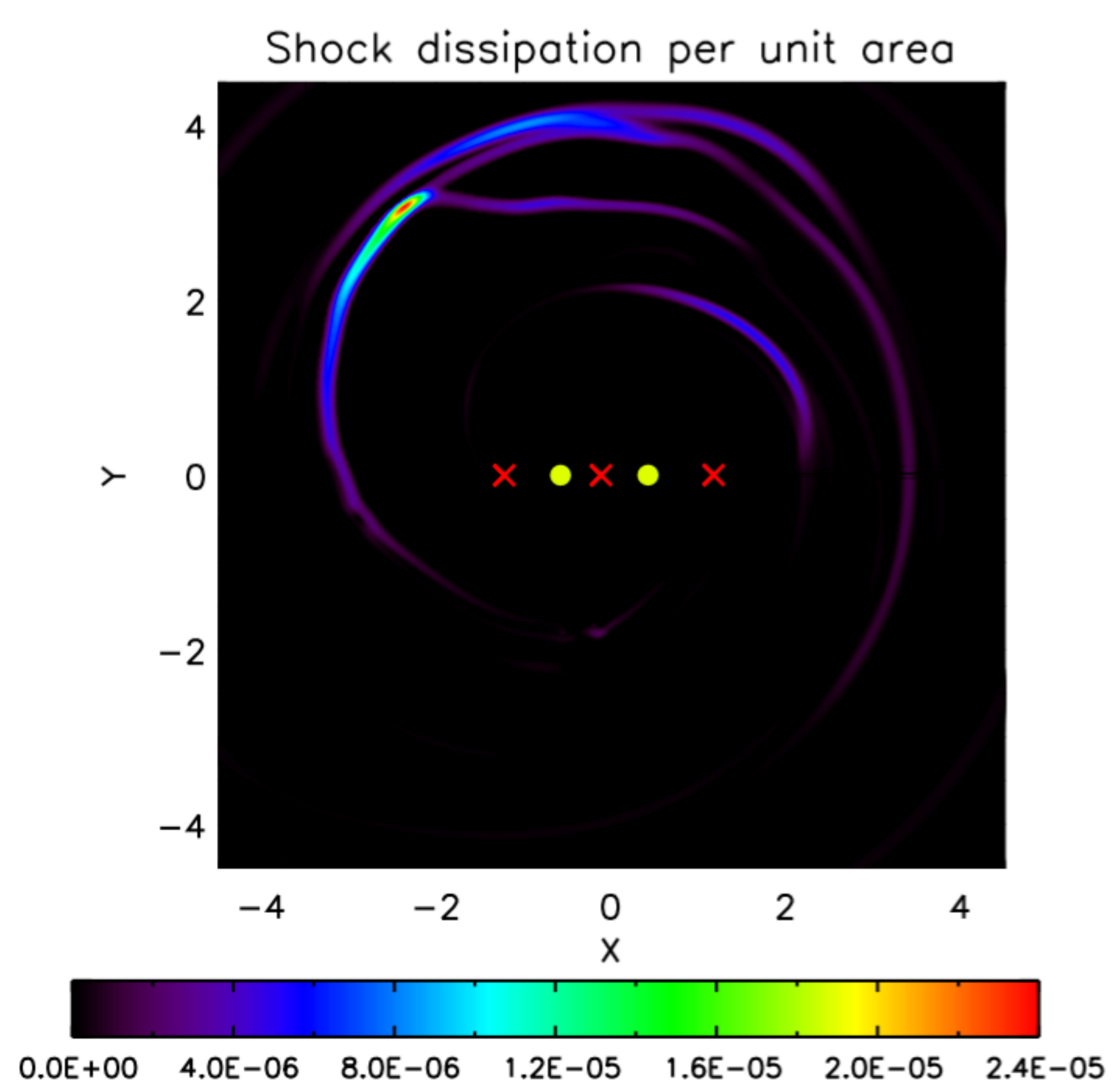}
\caption{ As in figure~(\ref{INCHHRVISC208}) but for model B after $210$
  binary orbits.}
\label{INCHVISC210}
\end{center}
\end{figure}

\begin{figure}
\begin{center}
%\vspace{-0cm}
%\hspace{-2cm}
\includegraphics[scale=0.31]{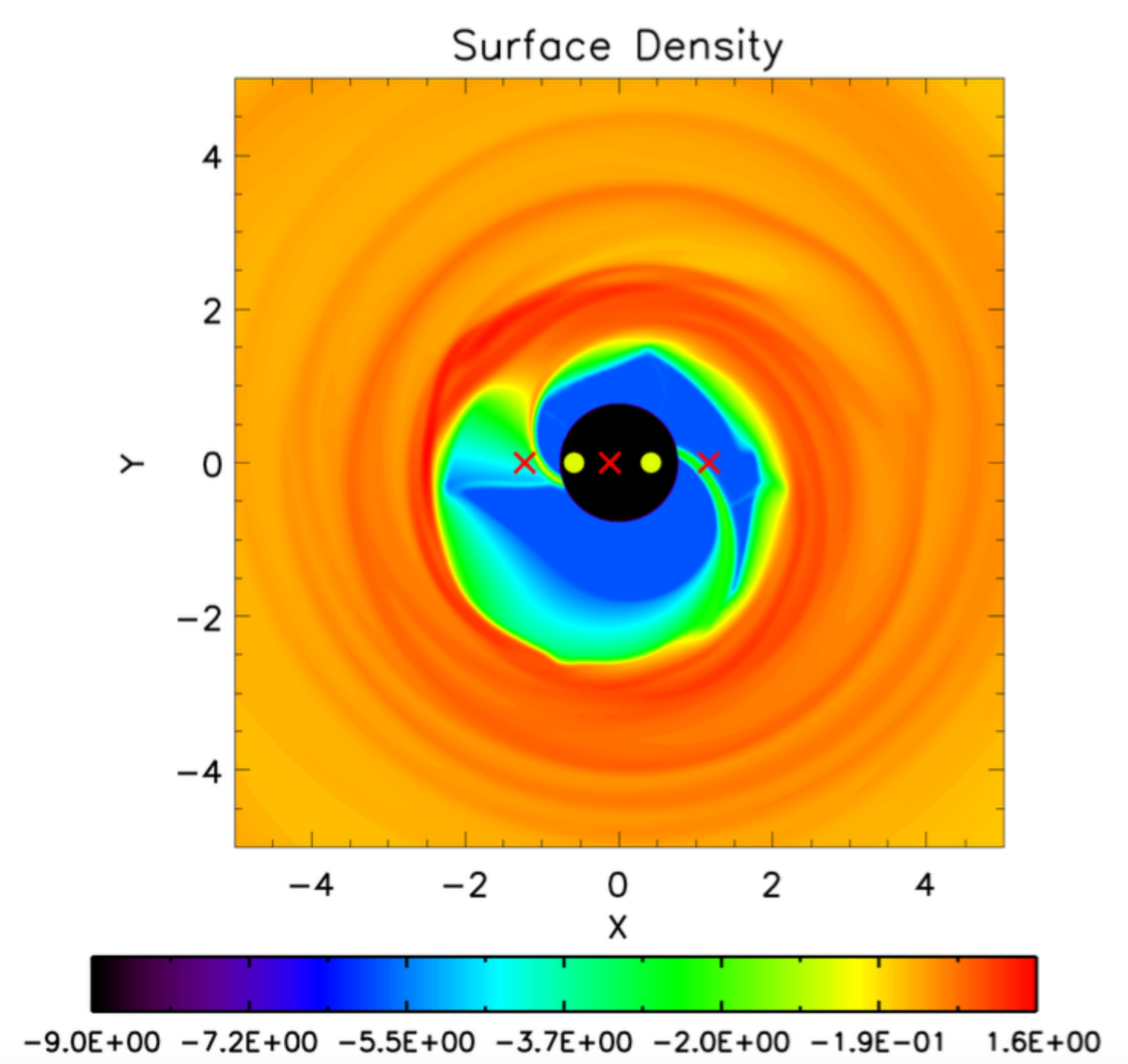}
%\vspace{3cm}\hspace{-0cm}
\includegraphics[scale=0.31]{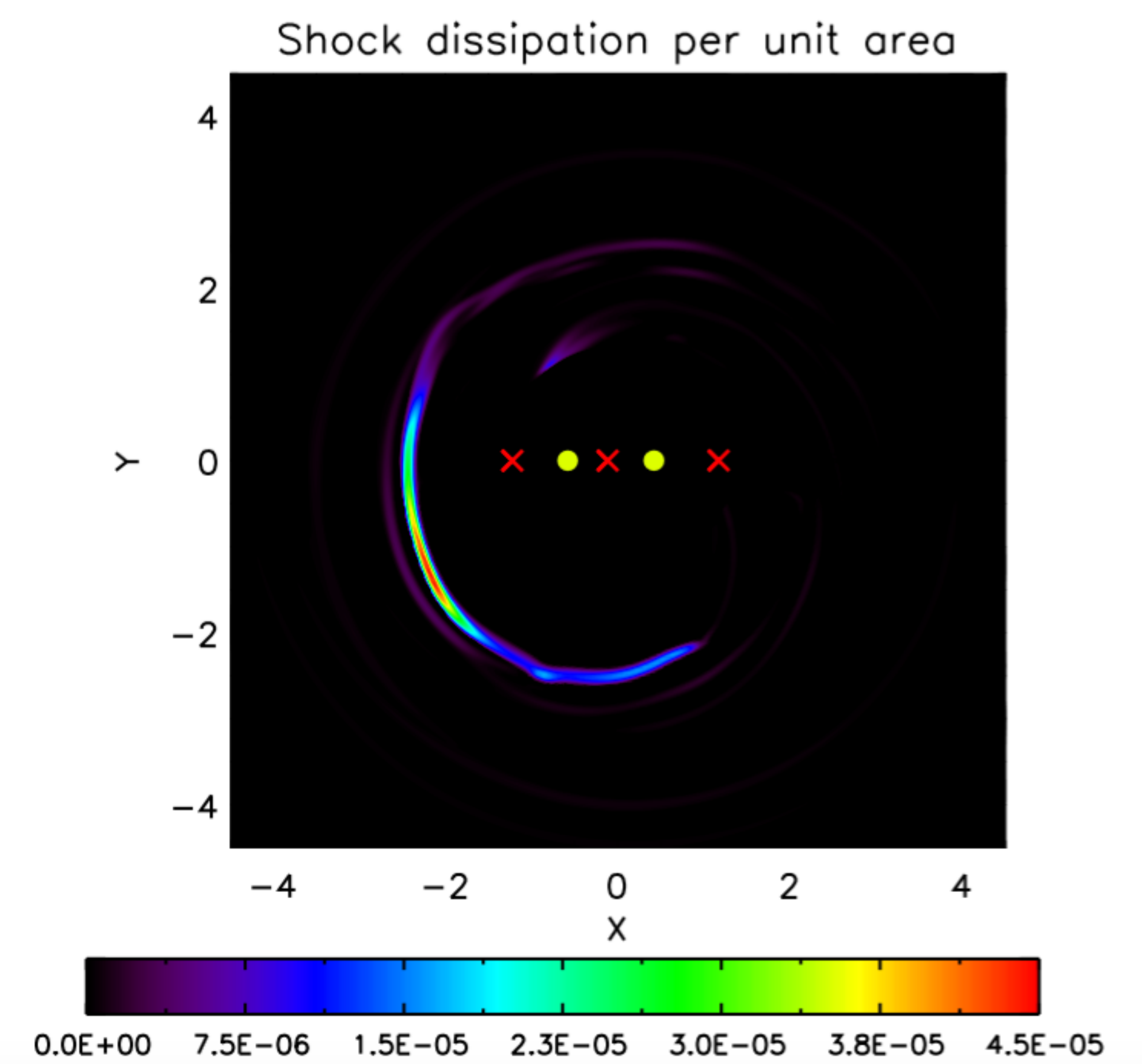}
\caption{ As in figure~(\ref{INCHHRVISC208}) but for model C after $208$
  binary orbits.}
\label{INCVINCVISC208}
\end{center}
\end{figure}

\begin{figure}
\begin{center}
%\vspace{-0cm}
%\hspace{-2cm}
\includegraphics[scale=0.31]{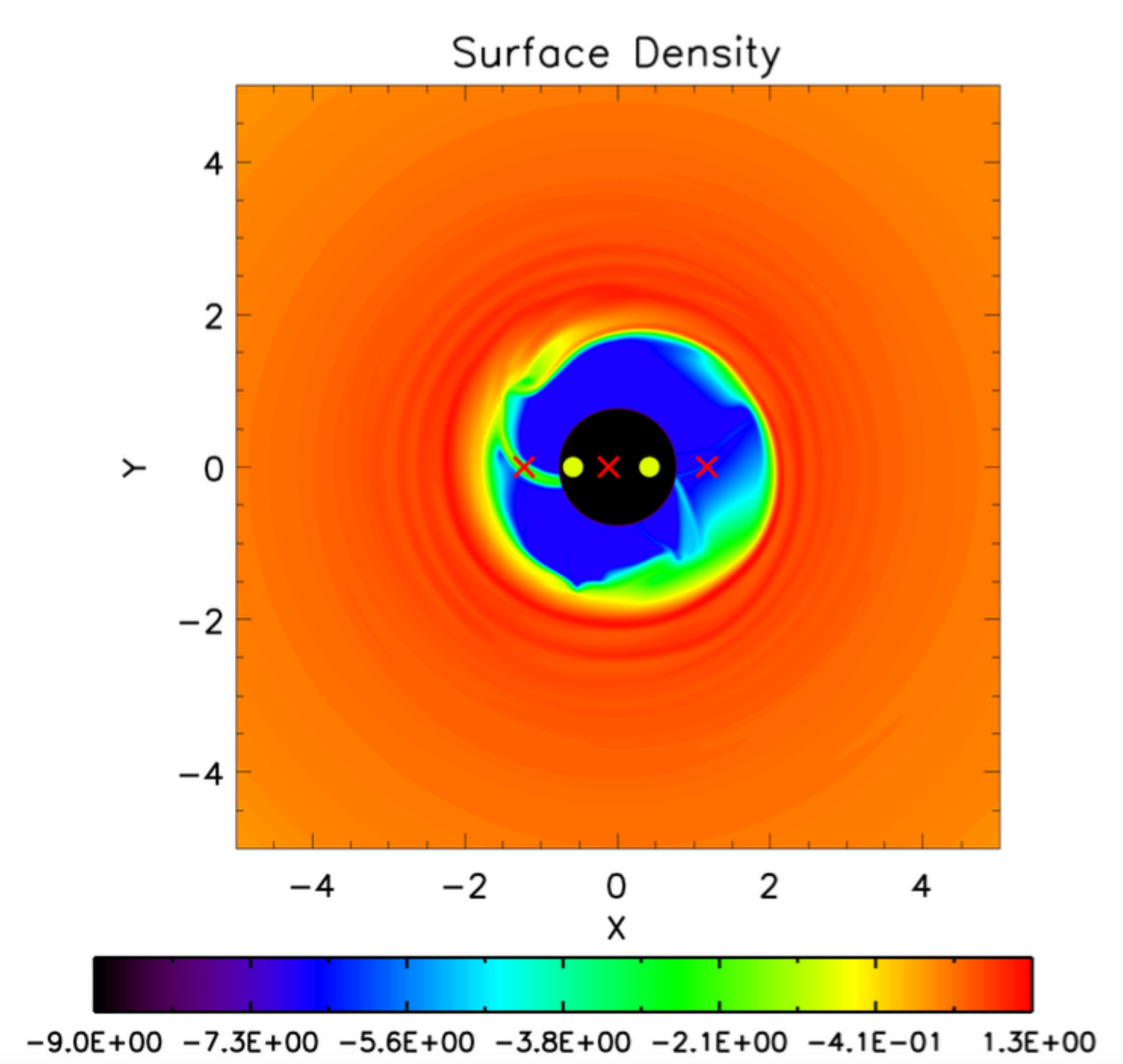}
%\vspace{3cm} \hspace{-0cm}
\includegraphics[scale=0.31]{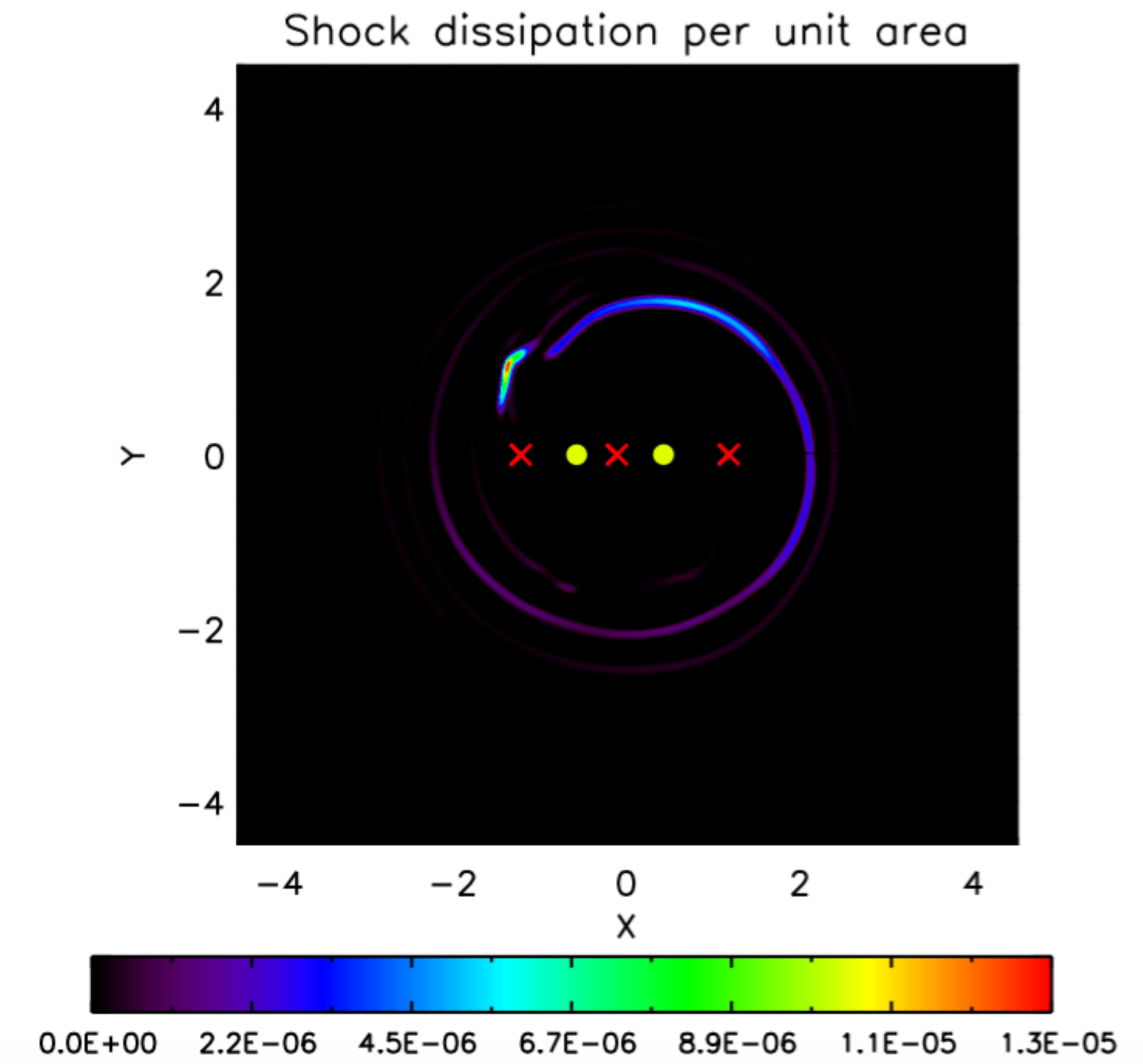}
\caption{ As in figure~(\ref{INCHHRVISC208}) but for model D after $208$
  binary orbits.}
\label{INCVISC208}
\end{center}
\end{figure}

Each of models $A$--$D$ was run for several hundred orbits.  After
about 50 orbits a cavity was formed in which the binary resided.  This
cavity has a radius between 1.5 and 3 and is in general not strictly
circular.  Inside this cavity, the surface density was dramatically
reduced.  However, accretion into the binary system occurs through
streams of material passing through the exterior Lagrangian points
$L_2$ and $L_3$.  These features are illustrated in surface density
contour plots for models A--D presented in
figures~(\ref{INCHHRVISC208})--(\ref{INCVISC208}).  These are shown
after $208$ binary orbits for models A, C and D and after 210 orbits
for model B. But note that the pattern looks similar at all times.
These figures also show the distribution of the rate of dissipation of
energy associated with shocks.  This is here traced by evaluating the
rate of energy dissipation associated with  artificial
viscosity. We see that this is highly localised in radius in regions
near the boundary of the cavity, with $ 1.5 < r <3$, as would be
expected for dissipation associated with tidal torques that maintain
the cavity.

\begin{figure}
\begin{center}
%\vspace{-5cm}\hspace{-1.3cm}
\includegraphics[scale=0.38]{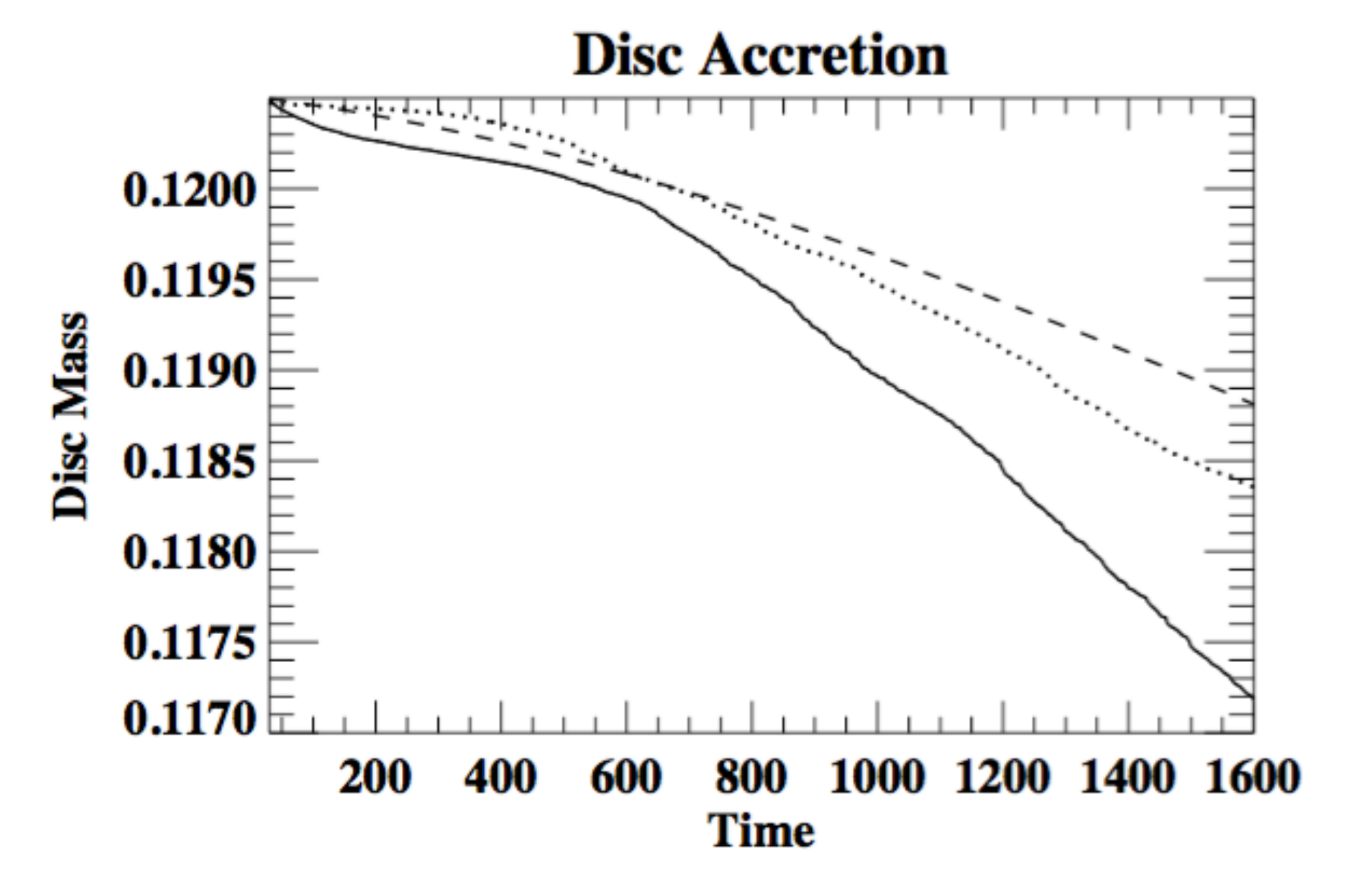}
%\hspace{-3.6cm}
\includegraphics[scale=0.38]{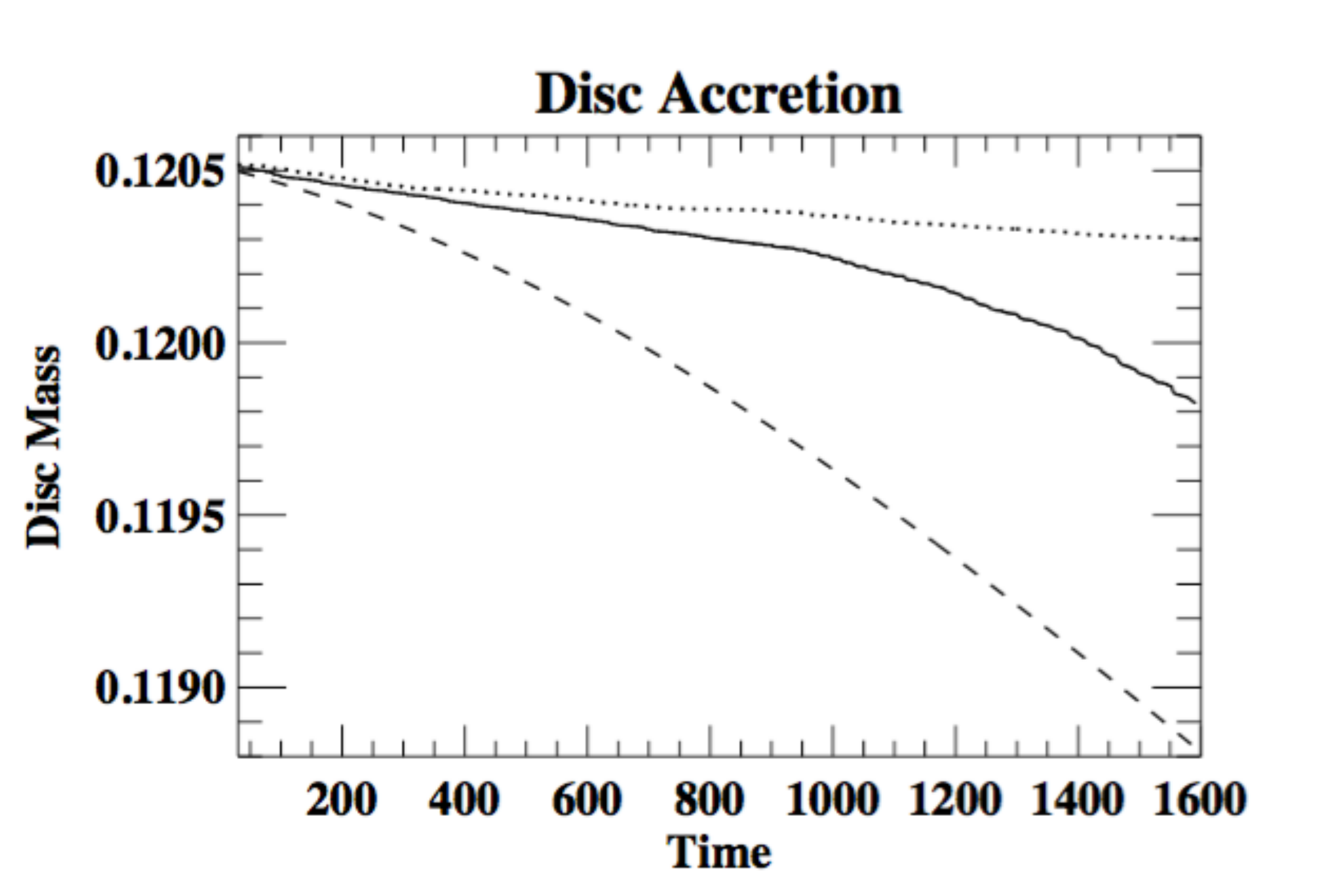}
\caption{ The upper panel shows the mass of the disc as a function
  of time in code units for model A ({\em solid line}), and model B
  ({\em dotted line}).  The {\em dashed line}, is for a disc with the
  same kinematic viscosity as these models but with the binary absent.
  The lower panel shows the mass of the disc as a function of
  time for model C ({\em solid line}) and model D ({\em dotted line}).
  The {\em dashed line} is for a disc with the same kinematic
  viscosity as model C but with the binary absent.  }
\label{figDiscMass}
\end{center}
\end{figure}

\noindent In figure~(\ref{figDiscMass}), we show the mass of the disc
as a function of time in code units for models A--D.  Also shown is
the evolution of the mass of a disc with the same kinematic viscosity
prescription as models A--C but with the binary absent. The slopes of
these curves give the accretion rate into the binary system. The
accretion rate averaged over time for $600 < t < 1600,$
$\langle{\dot M}\rangle,$ for models A--D, is given in
table~\ref{table2} in units of $\langle {\dot M}_0 \rangle ,$ the
latter being the average accretion rate for the disc without the
binary system.

\begin{table}
%\begin{centering}
\begin{tabular}{ccc}
\hline
 Model & $\langle{\dot M}\rangle/ \langle {\dot M}_0 \rangle $  \\
\hline
 A  & $2.0$ \\
 B  & $1.43$\\
 C  & $ 0.44$ \\  %0.26
 D  & $0.087$  \\
\hline
\end{tabular}
\caption{ Mean accretion rates for $ 600 < t < 1600,$ for the models 
  discussed in the text, expressed in terms of the mean accretion rate evaluated
  over the same time interval  for a disc
  in which the central binary was absent.
  Note that in units  of the fiducial value ${\dot M}_f = 3\pi\nu \Sigma_0$,
  with $\nu = \alpha (H/r)^2D^2\omega$ 
   evaluated for model A (see table \ref{table1}),
  $\langle {\dot M}_0\rangle= 1.63{\dot M}_f.$ 
\label{table2}}
\end{table}

\noindent For models A and B, this quantity is respectively $2.0$ and
$1.43$, indicating that accretion is if anything modestly enhanced by
the presence of the binary. This is in line with results given by Shi
\& Krolik (2015). For model~C, which has the same kinematic viscosity
prescription but the smallest $H/r,$ it is reduced to $0.44$,
indicating that accretion is somewhat suppressed by the presence of
the binary.  For model~D, which has a same value of $H/r$ as model~C
but a value of $\alpha$ that has been reduced by a factor of two, the
averaged accretion rate ratio is reduced to $0.087.$ This is a factor
of $\sim 5$ smaller than the value for model~C, as compared to an
expected reduction of a factor of two. This situation arises because
 at early times, measured in units of the characteristic viscous time of the model,
the accretion rates  are increasing,
  %as a function of time 
  taken together with the fact that  the characteristic viscous time 
%evolution 
is a factor of two longer  for model~D as compared to the
other models. If we compare the average accretion rate in model~C over
the time interval $ 300 < t <800,$ which contracts the time limits by
a factor of two, with the above rate for model~D evaluated as an
average for $600 < t <1600,$ it is smaller by a factor $\sim 2$, as
expected.  We remark that $\langle {\dot M}_0\rangle= 1.63{\dot M}_f,$
where the fiducial accretion rate ${\dot M}_f = 3\pi\nu \Sigma_0$ with
$\nu = \alpha (H/r)^2D^2\omega$
   evaluated for model~A.
  In addition, we note that a reduction in
the ratio $\langle {\dot M} \rangle/ \langle {\dot M}_0 \rangle $, 
similar to what we find here as $H/r$ is decreased, has been reported
in SPH simulations by Ragusa et al. (2016).  This corresponds to a
suppression of the accretion rate into the binary system as a result
of the action of tidal torques that is not seen in the simulations
with larger $H/r.$

However, it is important to note that in spite of this suppression, in
a steady state we expect that the mass accretion rate at large
distances should be the same as that onto the central binary.
Although it is impracticable to perform simulations for long enough
with an adequate dynamic range in such cases, we expect that when
suppression is large, the disc surface density just outside the cavity
edge will significantly exceed that at large distances and that in the
case of constant kinematic viscosity (see section~\ref{stdisc} below)
the ratio of these will be approximately equal to
$\langle {\dot M}_f \rangle/ \langle {\dot M} \rangle .$ When
considering the rate of energy dissipation due to tidal torques, as in
this paper, the value of the surface density just outside the cavity
edge, where these torques are applied, is the relevant one. This leads
to the fiducial rate of energy dissipation
$\langle {\dot M}_f\rangle D^2\omega^2$, rather than
$\langle {\dot M}\rangle D^2\omega^2$ (~see below). This can be
estimated without connecting conditions near the cavity edge to
conditions at large distance.

As indicated above, angular momentum is transferred from the binary
system to the disc through that action of compressional shocks and
viscous friction.  This transport maintains the cavity inside which
the binary resides. The associated dissipation occurs independently of
whether accretion into the binary system is suppressed or not.

\begin{figure}
\begin{center}
%\vspace{-7cm}
\includegraphics[scale=0.38]{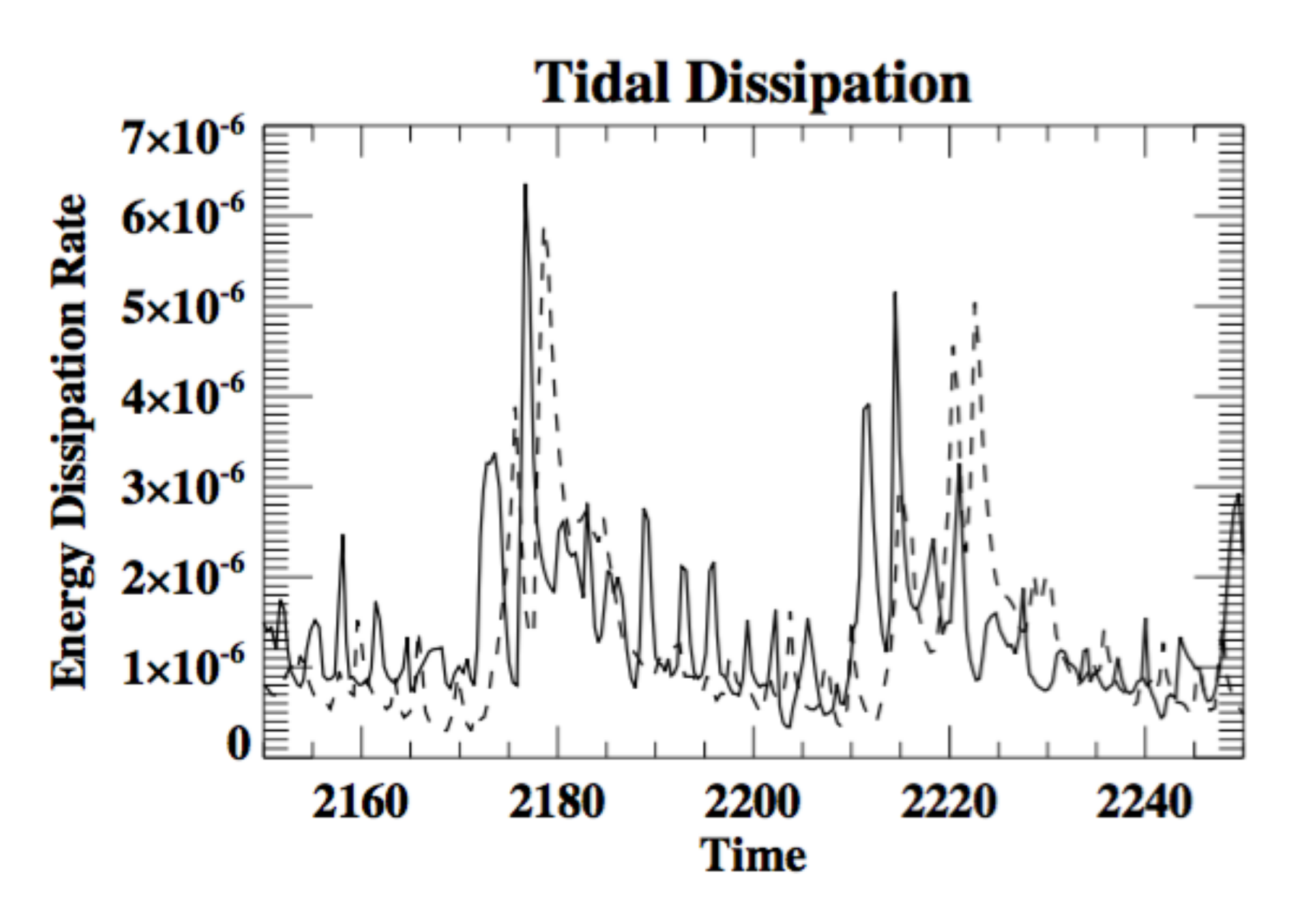}\\
%\vspace{-8cm}
\includegraphics[scale=0.38]{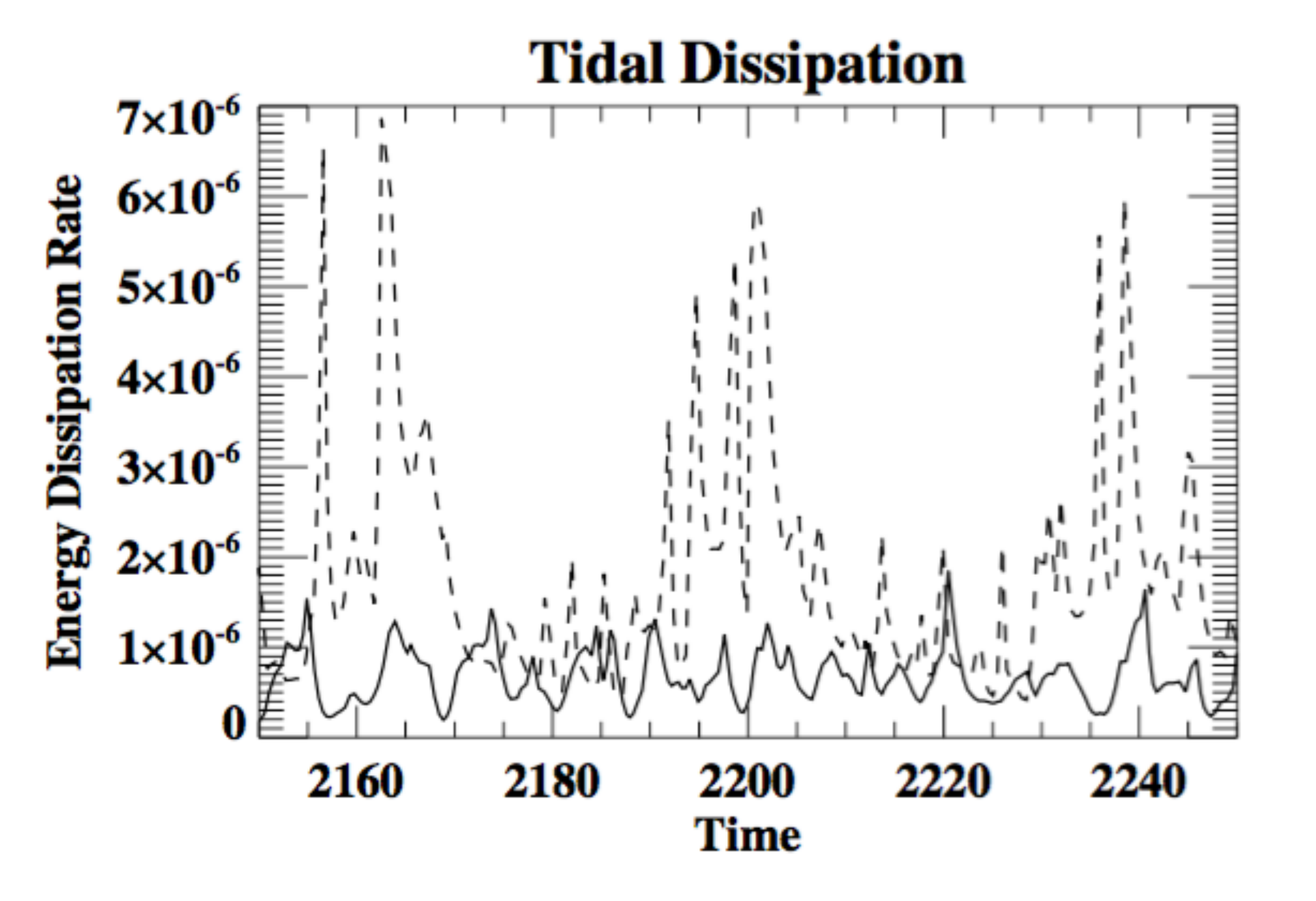}
\caption{ The  upper panel shows the total rate of shock
  dissipation as a function of time for $r > 1.6,$ in code units, for
  model~A ({\em solid line}) and model~B ({\em dashed line}). The
  lower  panel gives the corresponding plot for models~C ({\em
    solid line}) and~D ({\em dashed line}).  Note that, in code units,
  the fiducial rate of dissipation given by
  ${\dot M}_fD^2\omega^2 = 3\pi\nu \Sigma_0D^2\omega^2$, with
  $\nu = \alpha (H/r)^2D^2\omega$ 
     evaluated for model~A (see
  table \ref{table2}), is $8.5\times 10^{-7}.$ }
\label{shockD}
\end{center}
\end{figure}

The upper panel of figure~(\ref{shockD}) shows the total rate of shock
dissipation as a function of time for $r > 1.6,$ in code units, for
models~A and~B. The lower panel gives the corresponding quantity for
models~C and~D.  We note that the fiducial rate of dissipation
associated with tides, indicated here through the rate of energy
dissipation associated with artificial viscosity, is
$\epsilon_{f}= {\dot M}_fD^2\omega^2 = 3\pi\nu \Sigma_0D^2\omega^2$,
with $\nu = \alpha (H/r)^2D^2\omega$ evaluated for model~A, and is
equal to $8.5\times 10^{-7}.$ For models~A--C, the average dissipation
rate is $\sim 2\epsilon_{f}$ although most of this is concentrated in
localised spikes.  In model~D, for which the effective viscosity was a
factor of two smaller, the average rate of dissipation
$\sim 0.5\epsilon_{f}.$ These results indicate that, unlike the
accretion rate, the tidal dissipation is not greatly inhibited for
small $H/r.$ This occurs at approximately the same rate in all cases
and this rate is approximately
$\langle {\dot M}_f\rangle D^2 \omega^2.$ Note that, as indicated
above, the reduction by a factor of four rather than a factor of two
in the case of model~D as compared to model~C can be accounted for by
the increase in the rate of evolution that occurs as a function of
time for models A--C.
 
\noindent
% \textcolor{blue}
{Figure~(\ref{shockD}) indicates that the rate of shock
dissipation is variable on a timescale on the order of a few
$\omega^{-1}$.}

%%
%frac{D}{D t} \int_V \rho d^3{\bf r}&=&  0 .\label{contg} 
%\label{mass_eq}
%\end{eqnarray}

\section{The origin of tidal dissipation}\label{Tidanal}
\label{sec:analysis}

We noted above that the rate of energy dissipation associated with the
tidal interaction is approximately given by
$\langle {\dot M}_f\rangle D^2 \omega^2,$ 
independently of whether accretion onto the central binary is
significantly suppressed.  We now explore a simple picture of this
dissipation.

\subsection{Rate of change of the Jacobi invariant}

We begin by considering an infinitesimal fluid element of mass $m$
occupying a varying area $A$ with surface density $\Sigma,$ going on
to derive an equation governing the evolution of the Jacobi invariant
associated with it.  The constancy of $m$ is consistently implied by
the continuity equation~(\ref{contg}).  From equation~(\ref{mog}) we
obtain:

\begin{equation}
  m \frac{D{\bf v}}{Dt} = \frac{D}{D t} \int_A \Sigma  {\bf v} \; {\rm
    d}A =  - m \nab \Phi  + {\bf F}_T  + m {\bf F}. 
\label{mom_eq}
\end{equation}

\noindent  Here:
\begin{equation}
{\bf F}_T 
{= \int_A \left( \nab {\bcdot}  {\bf T} \right) \; {\rm d}A }=
\int_{\Gamma}  {\bf T} {\bcdot} \hat{\bf n} \; {\rm d} l,
\end{equation}

\noindent where the $i^{\rm th}$--component of
  ${\bf T} {\bcdot} \hat{\bf n}$ is $T_{ij} \hat{n}_j$, and the second
  integral is obtained using Green's theorem.  It is taken over the
boundary $\Gamma$ of the fluid element, with $\hat{\bf n}$ being the
unit vector normal to the boundary and pointing outward, and so it is
derived from the stresses due to exterior fluid. 
 Thus it could act
like a drag force per unit mass.  For the sake of brevity, we set
${\bf f}= {\bf F}+{\bf F}_T/m.$

Using the fact that for our set up the gravitational potential $\Phi$
is a function of $\varphi$ and $t$ through the combination
$\varphi - \omega t,$ we obtain an equation for the rate of change of
the Jacobi invariant $E - \omega J$, where $E$ is the kinetic plus
potential energy and $J$ is the angular momentum of the fluid element.
We begin by noting that we can obtain an equation for the rate of
change of the kinetic energy of a fluid element from
equation~(\ref{mog}) under the form:

\begin{equation}
  \frac{1}{2}\frac{D}{D t} \int_A  \Sigma v^2 \; {\rm d}A = -
  m{\bf v} {\bcdot} \nab \Phi +m {\bf v} {\bcdot}  {\bf f} -\epsilon,
\label{energy_eq}
\end{equation}

\noindent where
\begin{equation}
\epsilon = \int_A  \left( \nab {\bf v} : {\bf T} \right) \; {\rm d} A ,
\label{work_eq}
\end{equation}

\noindent with
$\nab {\bf v} : {\bf T} = T_{ij} (\partial v_i / \partial x_j )$,
represents the rate of doing work by internal stresses.  This contains
the rate of dissipation due to viscosity when that is present.  We may
also write equation~(\ref{energy_eq}) as:

\begin{equation}
  \frac{D E}{D t} = \frac{D}{D t} \int_A \Sigma \left (\frac{1}{2}
    v^2+
    \Phi \right) \; {\rm d}A  =   m\frac {\partial \Phi}{\partial t} +
  m {\bf v} {\bcdot}  {\bf f} - \epsilon .
\label{energy2_eq}
\end{equation}

\noindent From the $\varphi$--component of equation~(\ref{mom_eq}), we
obtain:
 
\begin{equation}
  \frac{DJ}{Dt} = m\frac{D( rv_{\varphi}) }{Dt} = - m\frac{\partial
    \Phi}{\partial \varphi} +
  mr{\bf f} {\bcdot} \varphat ,
\label{amom_eq}
\end{equation}

\noindent where $\varphat$ is the unit vector in the azimuthal
direction.  Multiplying equation~(\ref{amom_eq}) by $\omega$ and
subtracting the resulting equation from equation~(\ref{energy2_eq}) we
obtain:  
 
\begin{equation}
   \frac{D(E-\omega J)}{Dt} =   m \left( {\bf
       v} - r \omega \varphat \right) {\bcdot} {\bf f}-\epsilon \; . 
\label{Jac_eq}
\end{equation}

\noindent Note that this form applies in the inertial frame.

\subsection{Relation between energy dissipation and binary torque for
  an oscillating fluid element}

Suppose a fluid element near some inner disc edge undergoes cyclic
behaviour such that, starting from a circular orbit, it moves inwards
and as a result of interacting with the binary and surrounding fluid
it then returns to the same edge where it is circularized.  Then the
total change in Jacobi constant is zero and from
equation~(\ref{Jac_eq}) we have:
 
\begin{equation}
  \int \left( - m{\bf v} {\bcdot} {\bf f}+\epsilon \right) \; {\rm d}t = \int -mr\omega
  \varphat {\bcdot} {\bf f} \; {\rm d}t =\omega\int \langle {\cal T}\rangle dt .
  % &=&0.  int \left( -m({\bf
  %   v}-r\omega\mbox{\boldmath${\hat{\varphi}}$})\cdot {\bf
  %   f}-\epsilon\right)dt &=&\int \left(
  %   -mr\omega\mbox{\boldmath${\hat{\varphi}}$})\cdot {\bf f}t &=&0.
\label{Jacan_eq}
\end{equation}

\noindent Here the time integrals are over the duration of the
process. Equation (\ref{Jacan_eq}) states that the mean rate of energy
dissipation through the action of the force ${\bf f},$ as measured in
the inertial frame, together with the mean rate of energy dissipation
by internal stresses, is the product of the binary orbital frequency
and time average of the torque exerted by the binary,
$\langle {\cal T}\rangle,$ taken over the duration of the process.
That this is the time averaged torque exerted by the binary follows
from (\ref{amom_eq}).  As by assumption the fluid elements angular
momentum is not changed through this process, the torque is
communicated to material in contact.

\noindent As we know accretion into the binary system is ongoing, the
fluid element will subsequently move away from the edge into the
binary.  A similar discussion could be applied to such a final stage
if one knows the change in $E-\omega J,$ the Jacobi invariant, during
this process. As it is dynamical we shall assume this is small so that
equation~(\ref{Jacan_eq}) applies throughout over extended time periods.

\subsection{ A standard accretion disc model}\label{stdisc}

We now connect the above discussion to standard accretion disc
modelling that performs azimuthally averaging.  Within this framework,
we consider a standard steady accretion disc with an inner boundary at
which the above torque is applied by the binary and through which
there is a constant accretion rate ${\dot M}.$ Angular momentum
conservation gives:

\begin{equation}
  \frac{d}{dr}\left (| {\dot M}| r^2\Omega   +2\pi \nu\Sigma
    r^3\frac{d\Omega}{dr}\right) = -\langle{\cal T}\rangle_1,
\label{accnd}
\end{equation}

\noindent where $\Omega$ is the local angular velocity and $\langle{\cal T}\rangle_1$ is the
applied time averaged  torque per unit radius. If we suppose that the torque is
applied in a thin layer at an inner edge, integrating through this
layer we get on the exterior  $+$ side where $r=r_+$:

\begin{equation}
\left.2\pi \nu\Sigma r^3\frac{d\Omega}{dr}\right|_+  = -\langle{\cal T}\rangle,
\label{accnd1}
\end{equation}

\noindent where $\langle {\cal T}\rangle $ is again the time averaged
total applied torque and $|_+$ indicates evaluation at $r=r_+.$
However, integrating equation~(\ref{accnd}) in the main body of the
disc we have:

\begin{equation}
  | {\dot M}| r^2\Omega   +2\pi \nu\Sigma r^3\frac{d\Omega}{dr} =  {\rm constant} =  | {\dot M}|r^2\Omega |_+ -\langle{\cal T}\rangle.
\label{accnd2}
\end{equation}

\noindent This equation determines the $\Sigma$ profile and rate of
viscous energy dissipation as in a standard disc model.  Note that, as
compared to the standard solution with a zero torque applied at the
boundary, there is the additional $\langle{\cal T}\rangle$ term.  This
is responsible for an additional viscous dissipation rate in the disc
amounting to $\Omega |_+\langle{\cal T}\rangle.$ This is comparable to,
but significantly less than, the tidal dissipation rate estimated above
as $\omega\langle {\cal T}\rangle.$ This is because the dissipation
envisaged there cannot all be associated with a simple shear
viscosity.  The difference
$(\omega -~\Omega |_+ )\langle {\cal T}\rangle$ must represent the
dissipation rate associated with compressional shocks (Papaloizou \&
Pringle 1977).

In addition, equation~(\ref{accnd1}) implies that the rate of
dissipation due to tidal torques acting on a Keplerian disc is
$\omega \langle {\cal T} \rangle = \left.3\pi \nu\Sigma r^2\omega \Omega\right|_+.$
This is similar to the fiducial rate seen in the simulations and given
by:
\begin{equation}
  \epsilon_f = {\dot M}_fD^2\omega^2 = 3\pi\nu \Sigma_0D^2\omega^2,
\label{Edot}
\end{equation}
\noindent where $\Sigma_0$ should be thought of as
being the value of $\Sigma$ just outside the cavity edge and $\nu$ is taken to be $\alpha(H/r)^2D^2\omega.$

However when tidal torques are effective at restricting
  accretion, as noted above, in a steady state the surface density
  just exterior to the gap edge will significantly exceed the surface
  density at large distances.  For the simple example where $\nu$ is
  constant, equation (\ref{accnd2}) implies that at larger distances
  $\Sigma \rightarrow \Sigma_{\infty},$ where
 \begin{equation}
 3\pi \nu\Sigma _{\infty}=   | {\dot M}| .
\end{equation}
Then:
 \begin{equation}
 \frac{\Sigma_+}{\Sigma_{\infty}}=\frac {\langle{\cal T}\rangle} {| {\dot M}|r^2\Omega|_{+}}.
\end{equation}
Thus, in this case, the mean energy dissipation rate due to tidal
torques will exceed $|{\dot M}|\omega r^2\Omega|_{+}$ by a factor
approximately equal to the steady state accretion rate estimated
adopting the value of the surface density just exterior to the cavity
edge to the actual steady state accretion rate.

\section{Particle simulations}
\label{sec:particle_sim}

In order to gain more insight into the nature and magnitude of the
tidal dissipation occurring near the cavity boundary, we use a
particle code to simulate the accretion flow around the binary.
This is based on the fact that, apart from where
  shocks occur, the pressure force is small compared to the
  gravitational force (or, equivalently, the sound speed is small
  compared to the Keplerian velocity), which implies, to a good
  approximation, that the motion should be ballistic (e.g., Shi et
  al. 2012).

\subsection{Computational setup}

The particles are set down at some distance $r$ from the binary center
of mass $O$ and are initiated the local  Keplerian angular velocity
$\Omega_K=\left[ G( M_1+M_2) /r^3 \right ]^{1/2} $ in the inertial
frame centered on $O$.
%Here $G$ is the gravitational constant and
%$M_t=M_1+M_2$ where $M_1$ and $M_2$ are the masses of the stars.  
The
particles move under the gravitational potential  due to the stars.  To
force them to move towards the cavity, we add a drag force per unit mass
${\bf F}$:

\begin{equation}
{\bf F} = - \eta \left| {\bf a} \right| \frac{{\bf
    v}}{ \left| {\bf v} \right| },
\label{Fdrag}
\end{equation}

\noindent where $\eta \ll 1$ is a free parameter, and ${\bf a}$ is the
acceleration in the inertial frame centered on $O$.  This form of the
drag force ensures that it acts opposite to the velocity and, as the
acceleration results mainly from gravitational forces, is $\eta$ times
smaller than the gravitational force per unit mass.
% In the simulations presented hereafter, we have fixed
% $\eta=10^{-3}$.
In practice, we only apply ${\bf F}$ to
particles beyond the inner radius of the cavity.

\noindent This drag force mimics the action of disc viscosity,
modelled using the $\alpha$ prescription above, but more realistically
resulting from turbulent processes arising from the magnetorotational
instability.  In a realistic situation, this would enable particles to
lose angular momentum and get accreted onto the binary.
%Pressure effects are ignored, which is a good approximation within the
%cavity, as the density there is very small.  Previous hydrodynamical
%simulations have indeed confirmed that, as expected, the motion of the
%particles within the cavity is essentially ballistic, as gravitational
%forces dominate over pressure forces and magnetic forces when they are
%present.

\noindent The parameter $\eta$ defined through equation~(\ref{Fdrag})
can be approximated by $\eta \sim |v_r|/v_{\varphi}$.  In a standard
accretion disc, $|v_r| \sim \nu / r$ where $\nu = \alpha c_s H$ is the
kinematic viscosity, with $c_s$ being the sound speed.  Using
$c_s = (H/r) v_{\varphi}$, we obtain $\eta \sim \alpha (H/r)^2$, which
gives $\eta \simeq 10^{-4}$ for models~A, B and~C and
$\eta= 5 \times 10^{-5}$ for model~D.  However, at the disc's inner
edge, the length scale is given by $H$ rather than $r$, and the
magnitude of the radial velocity in low surface density regions
may be better approximated by $|v_r| \sim \nu / H$.  This would then
lead to $\eta \sim \alpha (H/r)$, i.e.  $\eta \simeq 10^{-3}$ for all
the models~A--D.  Given that there is some uncertainty in the value of
$\eta$ that should be used , we have varied $\eta$ between $10^{-5}$
and a few times $10^{-3}$.

Before being accreted by one of the stars, a particle may be
accelerated onto a trajectory that sends it back to the circumbinary
disc.  When that happens, we assume that shocks recircularize the
orbit of the particle and accordingly we reset its velocity to be the
Keplerian value.

%\noindent In the corotating frame with origin at the centre of mass,
%we note $(x,y,z)$ the cartesian coordinates, where $x$ is the axis
%passing through the centres of the stars and $z$ is perpendicular to
%the orbital plane.  The coordinate of the primary star, of mass $M_1$,
%is $x_1=-a M_2/ M_t$, whereas the coordinate of the secondary
%star, of mass $M_2$, is $x_2=a M_1/ M_t$, where $a$ is the
%separation.

\subsection{Numerical results}

\begin{figure}
\begin{center}
\vspace{-2.cm}
{\includegraphics[scale=0.42]{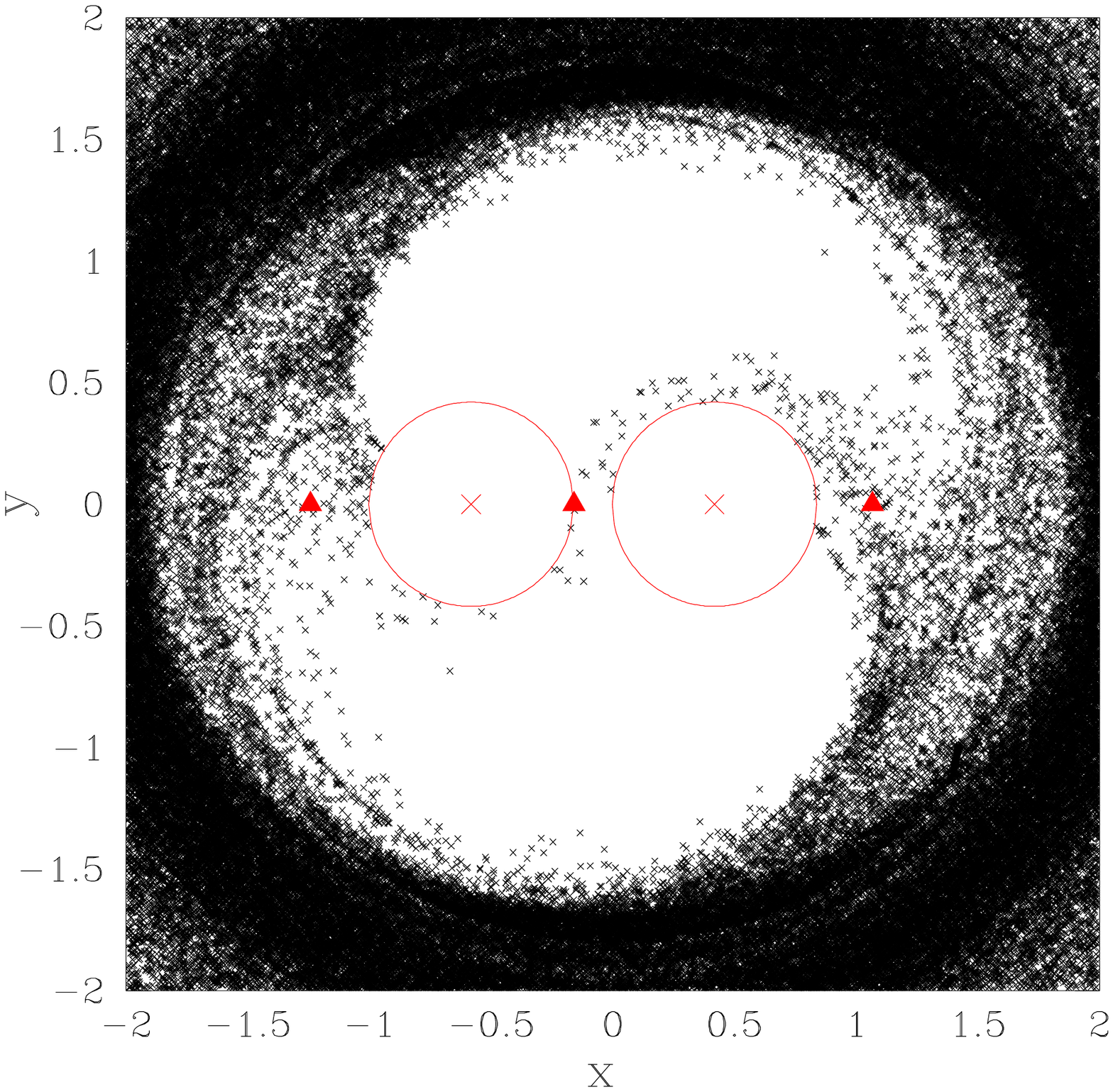}}
\vspace{-2.cm}
\caption{Accretion flow around a binary system in the corotating frame
  simulated with a particle code.  The mass of the primary (right red
  cross) is 7/12 and that of the secondary (left red cross) is 5/12.
  The separation is $D=1$.  The red circles indicate the outer edge of
  the circumstellar discs, which radius is about $0.4D$. The red
  triangles indicate the position of the Lagrangian points $L_3$, $L_1$
  and $L_2$ from right to left. }
\label{fig1}
\end{center}
\end{figure}

\noindent The accretion flow is shown in figure~(\ref{fig1}). 
As for the hydrodynamic simulations, we have $M_1=7/12$, $M_2=5/12$
and distances are given in units of the binary separation ($D=1$).
The particles were initially set up on circular orbits with $r$
between $2$ and $4.4$.  The small black crosses represent the position
of the particles and the large red crosses the position of the two
stars, with the primary on the right--hand side, as in
figures~(\ref{INCHHRVISC208})--(\ref{INCVISC208}).  The red triangles on
the figure indicate the position of the Lagrangian points $L_3$, $L_1$
and $L_2$ from right to left.  The value $\eta=10^{-3}$ was adopted
for the calculation shown in figure~(\ref{fig1}), but the accretion flow
is very similar for smaller values of $\eta$.  Although it takes
longer for particles to drift inwards when $\eta$ is reduced, the
overall structure of the flow is not affected.

\noindent The rotation of the binary is anti-clockwise.  Because the
binary is rotating faster than the particles, in the co--rotating
frame the motion of the particles is clockwise.  Particles that get
close enough to one of the stars are captured onto orbits around the
star, and circumstellar discs form.  Tidal truncation of these discs
by the other star limits their outer radii, which are shown as the red
circles on the figure.  We find here that the radius $r_d$ of these
discs is 0.4 times the binary separation, close to the value of one
third predicted by theoretical studies (Paczynski 1977, Papaloizou \&
Pringle 1977).  Particles within these discs have been removed from
the figure.
%The inner red circles, which radii are set to be $D/3$, delimit the
%regions from which particles are removed from the code, to avoid the
%timestep becoming to small.

As can be seen from the plot, the mean radius of the cavity cleared
out by the binary is about 1.9, close to the value of 2 expected from
theory (see Lubow \& Artymowicz 2000 and references therein).  The
details of the flow beyond this radius cannot be captured by the
particle code, as hydrodynamical effects become important there.

The flow shown in figure~(\ref{fig1}) is very similar to that obtained
from our hydrodynamical simulations (see, e.g., fig.~[\ref{INCVISC208}])
and also those of, e.g.,  Hanawa, Ochi and Ando (2010), who considered a
binary with a mass ratio similar to that adopted here.

\subsection{Collisions at the edge of the cavity}

\label{sec:collisions}

As has been noted in previous studies of circumbinary discs (Shi \&
Krolik 2015), material entering the inner cavity does not in general
fall directly onto one of the circumstellar discs.  A particle
originating from the edge of the cavity
% has some initial angular momentum, it
may be accelerated by the gravitational torque on a trajectory that
propels it back towards the circumbinary disc.  Shocks may then occur
that result in circularization of the trajectory, the excess kinetic
energy being released and radiated away.

\noindent We model this process using the particle code in the
following way.  After a particle has entered the cavity, which is
assumed to have a radius $r_{\rm in}$, if it is flung back out to the
disc, i.e. is found with $r \simeq r_{\rm in}$ again, its velocity is
reset to the Keplerian value at this radius.  The particle then
resumes drifting towards the cavity.  We note $\delta E_K$ the kinetic
energy which is released, and which is the difference between the
kinetic energy $E$ of the particle when it reaches $r_{\rm in}$ and
the kinetic energy $E_{\rm in}$ of the same particle undergoing
Keplerian motion there.  To evaluate $\delta E_K$, we need in
principle to know the value of $r_{\rm in}$.  However, if we vary this
radius between $1.5$ and $2.5$ for example, $E_{\rm in}$ is found to
vary only by a factor of 0.6.  Since $E$ itself varies only by a
factor of order unity when the particle goes from $r=1.5$ to $r=2.5$,
$\delta E_K$ does not depend very sensitively on $r_{\rm in}$.  In
addition, as can be seen from figure~(\ref{fig1}), the cavity is not
actually circular.  Therefore, it is not essential for the argument to
know $r_{\rm in}$ precisely.

\noindent If the distance between the particle and one of the stars
becomes smaller than the radius $r_d$ of the circumstellar discs, it
is then supposed to be accreted by the star.  We fix $r_d=0.4$.

\begin{figure*}
\begin{center}
\vspace{-1.cm}
{\includegraphics[scale=0.7]{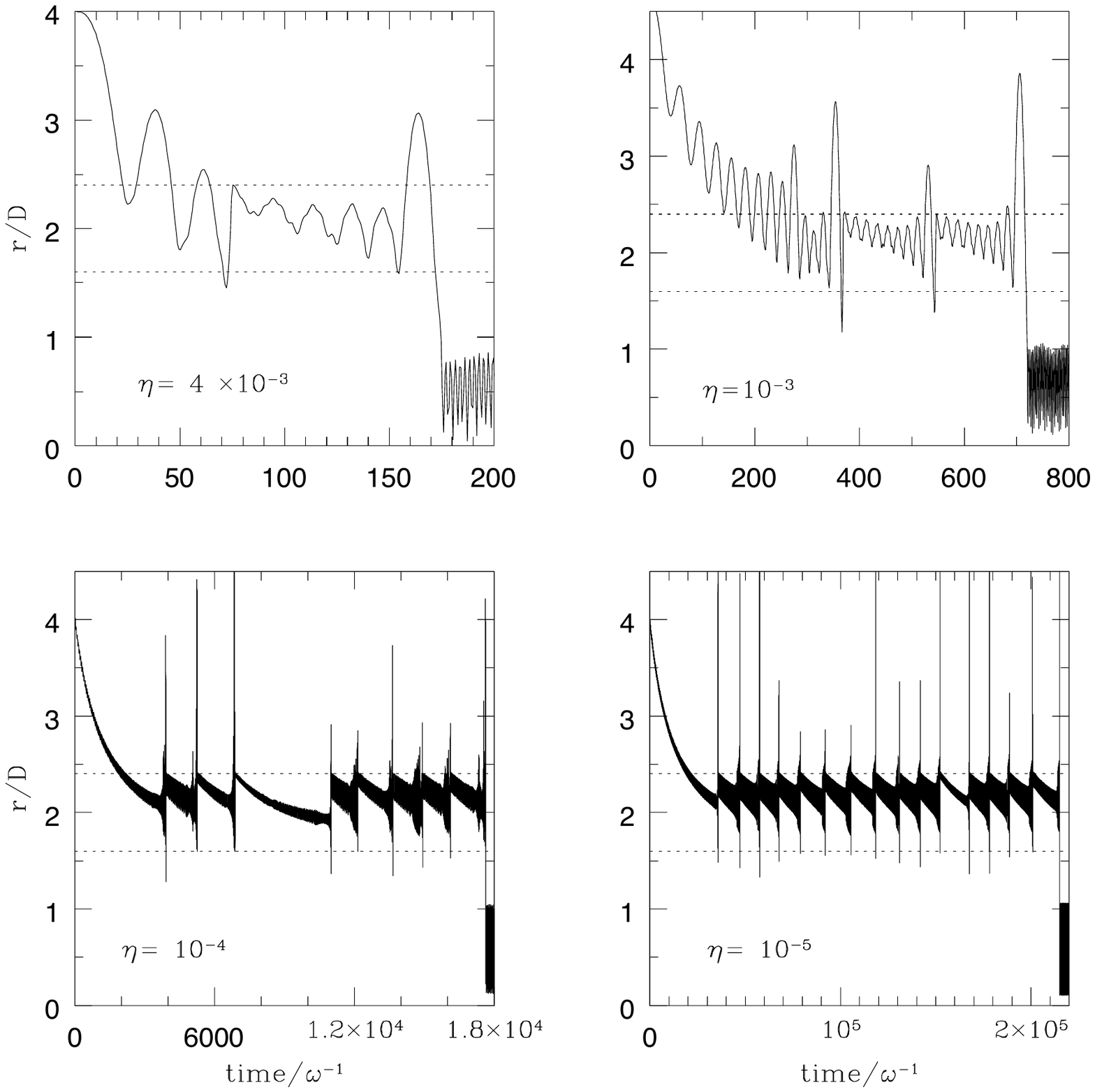} } 
\vspace{-4.cm}
\caption{ The radius $r$ (in units of the binary separation $D$) of a
  particle moving inside the cavity as a function of time (in units of
  $\omega^{-1}$) for $\eta=4 \times 10^{-3}$ ({\em upper left panel}),
  $10^{-3}$ ({\em upper right panel}), $10^{-4}$ ({\em lower left
    panel}) and $10^{-5}$ ({\em lower right panel}).  In this
  calculation, the velocity of the particle is reset to the Keplerian
  value if it is flung back to the disc and reaches $r=2.4$ after
  having gone down to $r=1.6$ (these values of $r$ are indicated by
  the {\em dotted lines}).  The trajectories that are shown are
  typical but note that trajectories may differ significantly from one
  particle to another.  As $\eta$ is reduced, the particle undergoes
  more collisions before crossing the binary orbit.}
\label{fig5}
\end{center}
\end{figure*}

The radius of a particle moving inside the cavity as a function of
time is shown in figure~(\ref{fig5}).  It moves slowly inwards in an
orbit with steadily increasing eccentricity.  It is then flung
outwards ultimately being reset on a circular orbit at larger radii
and the process then repeats until the particle is accreted by the
binary.  As can be seen, the particle undergoes both small and large
excursions, indicating that there should be a corresponding range of
shock strengths present as we found to be the case in our hydrodynamic
simulations (see fig.~[\ref{shockD}]).  In the calculations presented
in figure~(\ref{fig5}), the velocity of the particle is reset to the
Keplerian value if it is flung back to the disc and reaches $r=2.4$
after having gone down to $r=1.6$.  Therefore, circularization
proceeds only after a large excursion occurs.  Because we have chosen
$r=2.4$ rather arbitrarily as the radius where circularization takes
place, we cannot calculate the real time between two collisions with
this model.  However, we can calculate the number of times a particle
collides with the disc's inner edge before being accreted.  For a
fixed value of $\eta$, this number is obtained by averaging the
results corresponding to many trajectories calculated for different
initial conditions (for a given $\eta$, the trajectories for two
different particles may be significantly different, which is why an
averaging is required).

We find that, for $\eta=4 \times 10^{-3}$, on average particles get
flung back out to the disc having undergone one or two large radial
excursions.  For higher values of $\eta$, the drift force is rather
large and particles get accreted straight away without colliding with
the disc.  When collisions occur, particles hit the inner edge of the
circumbinary disc at $r \sim r_{\rm in}$ with a velocity which is on
the order of the Keplerian velocity at this radius but with a
different orientation.  Therefore, $\delta E_K \sim E_{\rm in}$.  If
the mass of particles drifting through the circumbinary disc's inner
edge per unit time is $\dot{M}$, then the total kinetic energy
released per unit time is
$ \Delta E_K \sim G (M_1+M_2) \dot{M} / r_{\rm in}$.  This is
essentially the same as given by equation~(\ref{Edot}) if we identify
$\dot{M} = \dot{M}_f ,$ the fiducial accretion rate which in steady
state is $3 \pi \nu \Sigma_0$ (see Section~\ref{setup} and
table~\ref{table2}), and it corresponds to what is found in our
hydrodynamic simulations in the cases of models~A and~B.

If $\eta$ is reduced to $10^{-4}$ or $10^{-5},$ particles move more
slowly inwards and undergo more large radial excursions leading to
more energy dissipated per particle as it is accreted.
Suppose this  is  ${\cal N}\delta E_K $ with ${\cal N}  > 1.$ In our
particle simulations, all the particles are  forced
inwards by the drag force given by equation~(\ref{Fdrag}), and
therefore they all end up being accreted by one of the stars.
Therefore, the  the total kinetic energy released per unit
time will be
$ \Delta E_K \sim {\cal N} G (M_1+M_2) \dot{M} / r_{\rm in}$.
 %accretion rate $\dot{M}$ onto the binary is the same
%for all values of $\eta$. 

\noindent However, we know from the more realistic hydrodynamic
simulations presented in section~\ref{sec:hydro_sim} and from the
results by Ragusa et al. (2016) that the accretion rate onto the
binary is actually reduced below the fiducial value ${\dot M}_f$ as
$H/r$ and $\alpha$ are reduced, which here corresponds to reducing
$\eta.$ On the other hand, as seen in the hydrodynamic simulations and
confirmed by the analysis done in section~\ref{sec:analysis}, the rate
of energy dissipation at the disc's inner edge is determined by the
tidal torque which maintains the cavity and is expected to be given by
$ \Delta E_K \sim G (M_1+M_2) \dot{M}_f / r_{\rm in}.$ This leads us
to identify ${\cal N} \sim \dot{M}_f /\dot{M}.$ In other words, the
total energy which is dissipated, being tapped from the binary,
depends only on the size of the cavity and the surface density at the
inner edge, not on the actual accretion rate onto the binary.

The above discussion implies that, when $H/r$ and $\alpha$ (or
equivalently $\eta$) are decreased, as particles are accreted more
slowly, each of them has to dissipate more energy, and therefore
undergo more collisions, which is exactly what is seen in the results
presented in figure~(\ref{fig5}).  Thus, in the case $\eta=10^{-4}$ or
$10^{-5}$, the tidal torques and related dissipation rate required to
maintain the cavity ( see eq.~[\ref{Edot}]) can be maintained with an
accretion rate onto the binary that is significantly less than the
fiducial rate ${\dot M}_f.$ This would correspond to models~C and~D
considered above where the accretion rate onto the binary is
significantly less than ${\dot M}_f.$

\begin{figure}
\begin{center}
%\vspace{-2.cm}
\hspace*{-1.cm} \includegraphics[scale=0.45]{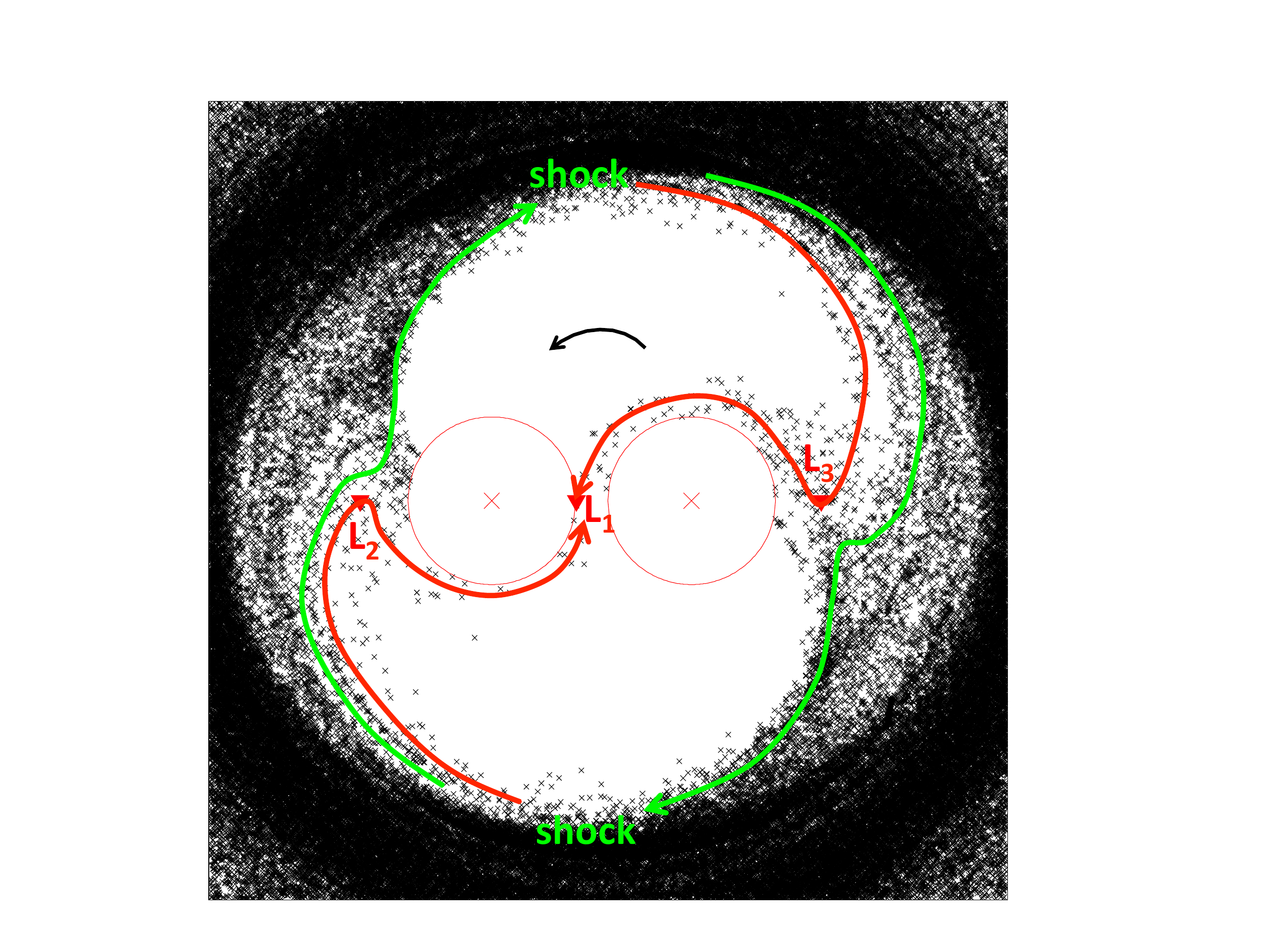}
%\vspace{-3.cm}
\caption{Sketch of the trajectories of the particles in the frame
  corotating with the binary near the edge of the circumbinary disc.
  The black arrow indicates the (anti--clockwise) direction of the
  binary's rotation.  The green lines show trajectories of particles
  which drift inwards but are propelled back to the disc and collide
  at the inner edge.  The red lines show trajectories of particles
  which go through $L_2$ or $L_3$.  They arrive at either of these
  points with a small  speed and are subsequently accelerated on an
  anti--clockwise trajectory around one of the stars.  }
\label{fig4}
\end{center}
\end{figure}

Figure~(\ref{fig4}) shows a sketch of the trajectories of the
particles in the frame corotating with the binary near the edge of the
circumbinary disc.  Particles that make it through one of the
Lagrangian points  enter into  orbital motion around one of the stars,
whereas other particles are propelled back to the disc after having
drifted inside the cavity.  As can be seen from this figure, the
shocks tend to be localized at some specific locations rather than be
uniformly distributed in azimuth.  Similar features were observed in
the hydrodynamic simulations presented in section~\ref{sec:hydro_sim},
although the location of the shocks  changes with time.

%
%===========================================================
%

\section{Disc's spectral energy distribution}

\label{sec:SED}

\begin{figure}
\begin{center}
{\includegraphics[scale=0.3]{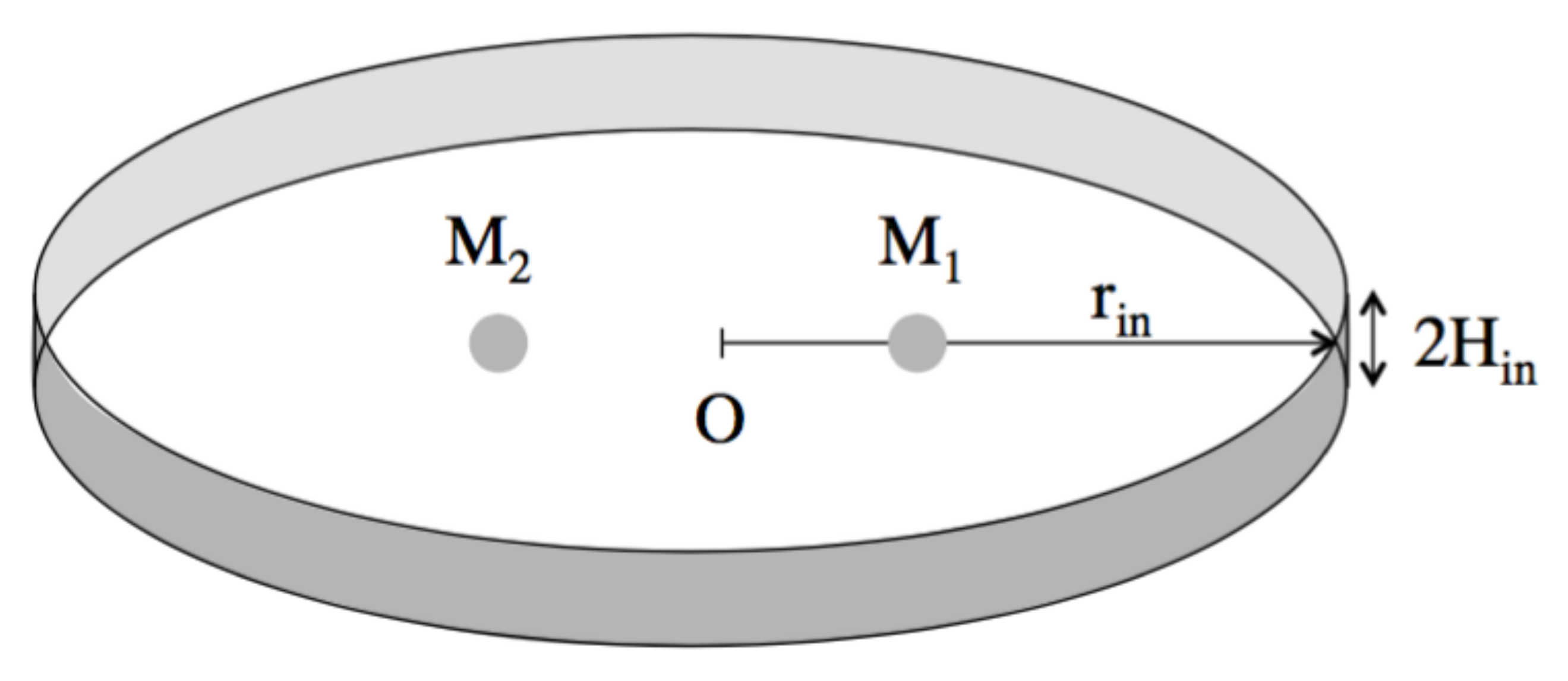}}
\caption{Schematic view of the circumbinary disc's inner edge and
  cavity. }
\label{fig2}
\end{center}
\end{figure}

In this section, we calculate the contribution from the collisions
described above to the disc's spectral energy distribution.  We model
the circumbinary disc's inner edge as a ``wall'' with radius
$r_{\rm in}$ and vertical extension $2 H_{\rm in}$, as illustrated in
figure~(\ref{fig2}).

\noindent At radii $r>r_{\rm in}$, the disc is heated by
irradiation from the central stars and by dissipation of gravitational
energy due to accretion.  These sources contribute the energies per
unit time per unit surface area $\sigma T_{\rm irr}^4$ and
$\sigma T_{\rm acc}^4$, respectively, where $\sigma$ is the
Stefan-–Boltzmann constant and (Friedjung 1985, Lynden--Bell \&
Pringle 1974):
\begin{eqnarray}
T_{\rm irr}^4 & = & \frac{2T_{\star}^4}{3 \pi } \left(  \frac{R_{\star}}{r}  \right)^3,
\label{Tdiscirr}
\\
T_{\rm acc}^4 & = & \frac{3 G M_{\star} \dot{M}_f} {8 \pi \sigma r^3}.
\label{Tdiscacc}
\end{eqnarray}

\noindent In equation~(\ref{Tdiscacc}), 
$\dot{M}_f$ is what has been defined in
section~\ref{sec:hydro_sim} as the {\em fiducial} accretion rate,
which in steady state is $3 \pi \nu \Sigma$.  
Here we have assumed $r \gg R_{\star}$ and replaced the two
stars with luminosities, radii and masses$L_{1,2}$, $R_{1,2}$ and
$M_{1,2}$, respectively, by one ``equivalent'' star located at the
center of mass with temperature $T_{\star}$ given by
$\sigma T_{\star}^4=L_1/(4 \pi R_1^2) + L_2/(4 \pi R_2^2)$, radius
$R_{\star}$ such that
$\sigma T_{\star}^4=(L_1+L_2)/(4 \pi R_{\star}^2)$ and mass
$M_{\star}=M_1 +M_2$.  Therefore, the surface temperature
$T_{\rm disc}$ of the disc at $r>r_{\rm in} $ can be calculated from:
\begin{equation}
T_{\rm disc} = \left( T_{\rm irr}^4 + T_{\rm acc}^4 \right)^{1/4}.
\end{equation}

\noindent Note that equation~(\ref{Tdiscirr}) assumes that the disc is
of constant aspect ratio, i.e. $H/r$ is constant.  If the disc were flared, the
temperature due to irradiation would be larger at larger radii, but
since we are only concerned here with the spectral energy distribution
near the disc's inner edge we will ignore such complications.  We have
also assumed in using equation~(\ref{Tdiscirr}) that the entire star
could be seen from any point at the surface of the disc.  This would not
be the case if $H_{\rm in}$ were comparable to or larger than $R_{\star}$, in
which case the temperature due to irradiation would be significantly
smaller than that given by equation~(\ref{Tdiscirr}).  As this would
only reinforce the conclusions presented below, we will also ignore
this effect.

At $r=r_{\rm in}$, the disc's inner wall is heated by irradiation from
the central stars and by the energy released through the action of
tidal torques.  That release occurs through shock dissipation and
viscous friction in the hydrodynamic model and collisions that result
in circularization just beyond the disc edge in the particle model.
These sources, respectively, contribute the energies per unit time per
unit surface area $\sigma T_{\rm in, irr}^4$ and
$\sigma T_{\rm in,coll}^4$, with:
\begin{eqnarray}
T_{\rm in, irr}^4 & = &\frac{L_1+L_2}{4 \sigma \pi r_{\rm in}^2},
\label{Tinirr}
\\
T_{\rm in, coll}^4 & = & \frac{\Delta E_K}{\sigma S_{\rm in}},
\label{Tincoll}
\end{eqnarray}
where $S_{\rm in} = 4 \pi f r_{\rm in} H_{\rm in}$ is the surface of
the wall over which collisions occur.  The factor $f<1$ accounts for
the fact that shocks are not distributed uniformly in azimuth.  From
the point of view of the particle model, we have assumed that the
particles that are propelled back to the disc collide and are
thermalized at $r=r_{\rm in}$.  This is consistent with the fact that
the hydrodynamic simulations presented in section~\ref{sec:hydro_sim}
(see also Shi \& Krolik~2015) show that the edge of the cavity is very
sharp, with the mass density dropping by several orders of magnitude
over a few scale heights.  Therefore, the particles that drift in the
cavity and are propelled back to the disc do not penetrate
significantly through the disc.

\noindent At locations where shocks occur, i.e. over a
surface of area  $S_{\rm in}$, the temperature $T_{\rm in}$ at the disc's inner
edge can be calculated from:
\begin{equation}
T_{\rm in} = \left( T_{\rm in,irr}^4 + T_{\rm in,coll}^4 \right)^{1/4},
\end{equation}
whereas at locations where there are no shocks, i.e. over a surface area
$(1-f) S_{\rm in}/f$ ,  it is given by $T_{\rm in, irr}$.

\noindent  Hereafter we will take:
\begin{equation}
\Delta E_K = \frac{G(M_1+M_2) \dot{M}_f}{r_{\rm in}},
\end{equation}
as this gives the correct order of magnitude for the energy released
by the collisions (see previous sections and eq.~[\ref{Edot}]).  Here
again, $\dot{M}_f$ is the {\em fiducial} accretion rate, as this is
what sets the level of energy dissipation at the disc inner edge, even
when the accretion rate onto the binary is reduced.

The flux received from the disc at a wavelength $\lambda$ is then
$\lambda F_\lambda = (\cos i / L^2) E_\lambda$, with:

\begin{eqnarray}
E_\lambda & = & \int_{r_{\rm in}}^{r_{\rm disc}} \lambda B_{\lambda} \left[
  T_{\rm disc} (r) \right] 2 \pi r dr \nonumber \\  & + & 
4 \pi  r_{\rm in} H_{\rm in} \lambda \left[ 
f B_{\lambda} \left(
  T_{\rm in} \right)  + (1-f)  B_{\lambda} \left(
  T_{\rm in, irr} \right) \right],
\label{Elambda}
\end{eqnarray}

\noindent where $B_{\lambda}$ is the Planck function, $L$ is the
distance to the binary system and $i$ is the angle between the light
of sight and the perpendicular to the orbital plane.  In the
calculations below, we will adopt $r_{\rm in}=2D$,
$H_{\rm in}=0.05 r_{\rm in}$ and $r_{\rm disc}=10^3 r_{\rm in}$ (the
value of $r_{\rm disc}$ is not important as long as it is large enough
to capture all the contribution to the flux at the wavelengths we are
interested in).

\begin{figure*}
\begin{center}
\vspace{-2.cm}
{\includegraphics[scale=0.8]{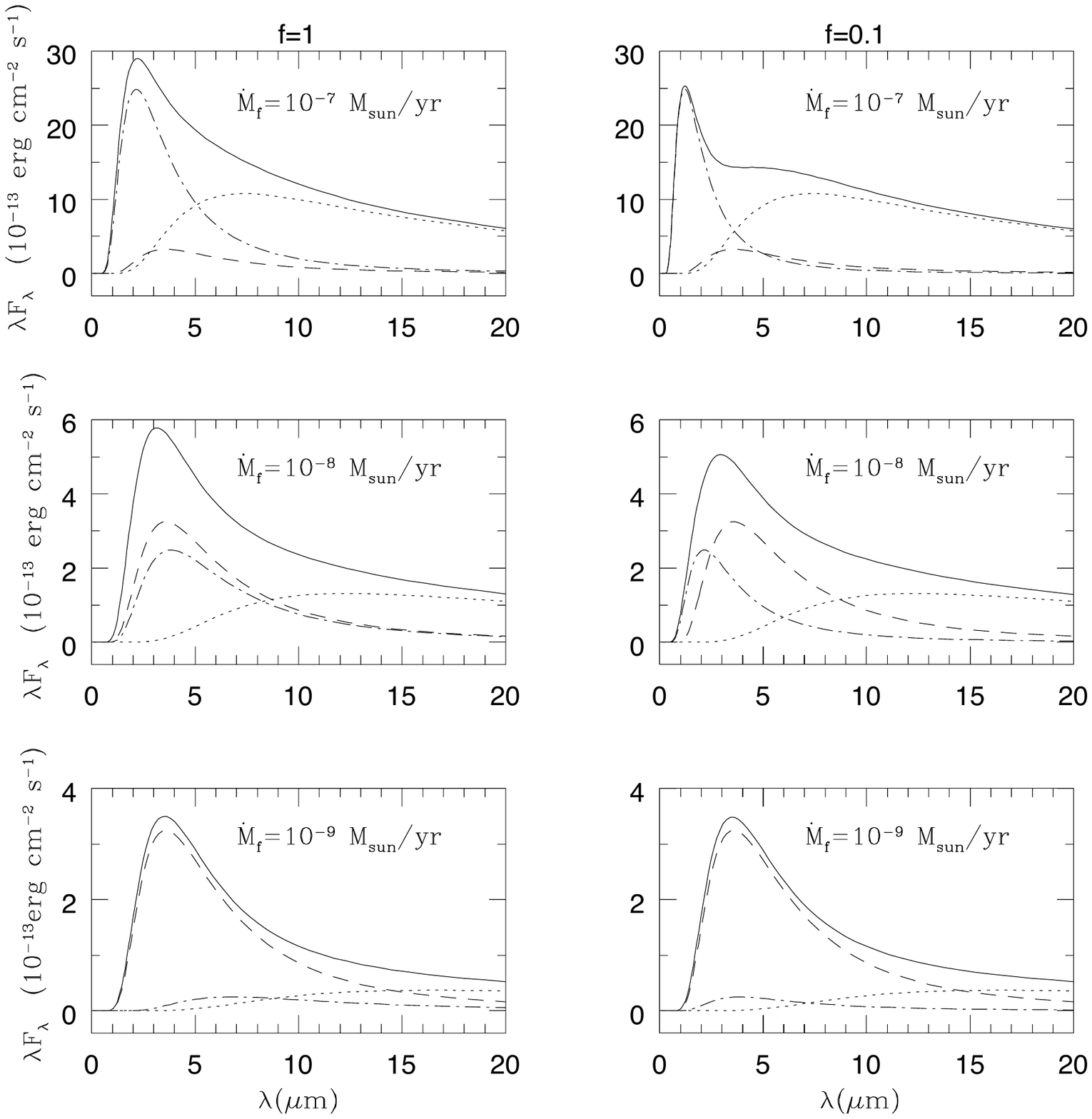} \vspace*{-2.cm}}
\vspace{-3.cm}
\caption{ Flux $\lambda F_{\lambda}$ (in units $10^{-13}$
  ~erg~cm$^{-2}$~s$^{-1}$) as a function
  of $\lambda$ (in $\mu$m). The {\em dotted lines} represent the disc
  without inner wall, i.e. only the integral is taken into account in
  calculating $E_{\lambda}$ in eq.~(\ref{Elambda}).  The {\em dashed
    lines} show the contribution of the inner wall when only
  irradiation is present, i.e. eq.~(\ref{Elambda}) without the
  integral and $T_{\rm in} = T_{\rm in, irr}$.  The {\em dot--dashed
    lines} show the contribution of the inner wall when only
  tidal torques  are present, i.e. eq.~(\ref{Elambda}) with only the
  $f B_{\lambda} \left( T_{\rm in} \right)$ term and
  $T_{\rm in} = T_{\rm in, coll}$.  The {\em solid lines} represent
  the total flux, i.e. disc and inner wall with both irradiation and
  tidal torques.  The plots on the {\em left--} and {\em right--hand sides} correspond to
  $f=1$ and $f=0.1$, respectively.  The {\em upper}, {\em middle} and
  {\em lower plots} correspond to $\dot{M}_f=10^{-7}$, $10^{-8}$ and
  $10^{-9}$~$M_{\sun}$~yr$^{-1}$, respectively.  The curves correspond
  to the parameters of the eclipsing binary CoRoT~223992193 (Gillen et
  al. 2014): $L_1=0.275$~L$_{\sun}$, $L_2=0.182$~L$_{\sun}$,
  $R_1=1.3$~R$_{\sun}$, $R_2=1.1$~R$_{\sun}$, $M_1=0.7$~M$_{\sun}$,
  $M_2=0.5$~M$_{\sun}$, $D=0.05$~au, $i=85^{\circ}$ and
  $L=756$~pc.}
\label{fig3}
\end{center}
\end{figure*}

Figure~(\ref{fig3}) shows the flux $\lambda F_\lambda$ as a function
of $\lambda$ for the parameters derived for the eclipsing binary
CoRoT~223992193 (Gillen et al. 2014): $L_1=0.275$~L$_{\sun}$,
$L_2=0.182$~L$_{\sun}$, $R_1=1.3$~R$_{\sun}$, $R_2=1.1$~R$_{\sun}$,
$M_1=0.7$~M$_{\sun}$, $M_2=0.5$~M$_{\sun}$, $D=0.05$~au,
$i=85^{\circ}$ and $L=756$~pc. Curves have been calculated for
$\dot{M}_f=10^{-7}$, $10^{-8}$ and $10^{-9}$~$M_{\sun}$~yr$^{-1}$ and
for $f=1$ and $f=0.1$. As pointed out in
section~\ref{sec:particle_sim}, $f$ is smaller than 1, but we have
done a calculation with $f=1$ to show that the results do not depend
very sensitively on the value of $f$. This is because, although
$T_{\rm in, coll}$ given by equation~(\ref{Tincoll}) increases
slightly when $f$ decreases (as $f^{-1/4}$), the surface that
re--radiates the energy dissipated through shocks is reduced. The
net effect is therefore not very significant.

\noindent Figure~(\ref{fig3}) shows that at wavelengths between 1--4
and 10~$\mu$m, the spectral energy distribution (SED) is completely
dominated by emission from the disc's inner edge. This emission is
itself dominated by the shocks produced by  fluid  particles being propelled back
to the disc for  large fiducial accretion rates  exceeding about
$10^{-8}$~$M_{\sun}$~yr$^{-1}$.  The maximum of the
excess moves towards smaller wavelengths when the surface over which
shocks occur is decreased.  For smaller accretion rates, irradiation
by the stars is the main source of emission from the disc's inner
edge. For $\dot{M}_f<10^{-9}$~$M_{\sun}$~yr$^{-1}$, the SED is
indistinguishable from that corresponding to
$\dot{M}_f=10^{-9}$~$M_{\sun}$~yr$^{-1}$, as the contribution to the
SED from accretion  becomes  negligible.

%
%===========================================================
%

\section{Summary and discussion}
\label{sec:discussion}

In this paper, we have shown that energy has to be dissipated in the
circumbinary disc for the tidal torque to maintain the cavity.  Energy
dissipation happens at the inner edge of the disc through
compressional shocks produced by fluid that is propelled back to the
disc after drifting inwards into the cavity.  The rate of energy
dissipation does not depend on the actual accretion rate onto the
binary, only on what we call the {\em fiducial} accretion rate through
the disc, $\dot{M}_f$, which is determined by the state variables in
the disc just beyond the cavity edge.  This would be the accretion
rate onto the central object if it was a single star rather than a
binary and the material near the cavity edge was part of a putative
exterior steady state disc.  Energy dissipation occurs at the fiducial
rate even when the accretion rate onto the binary is reduced below
$\dot{M}_f$ (which happens when the aspect ratio and viscosity of the
disc is reduced).  The rate of energy dissipation is typically
$G (M_1+M_2) \dot{M}_f/r_{\rm in}$, where $r_{\rm in}$ is the inner
radius of the circumbinary disc.

Hydrodynamic simulations (see also the MHD simulations performed by
Shi \& Krolik 2015) show that the edge of the circumbinary disc is
very sharp.  Therefore emission of the energy released by the shocks
is localised at the edge.  In addition, this edge is irradiated by the
central stars and this also contributes to the SED of the system.

In a tight PMS binary with a separation of $\sim 10$~R$_{\sun}$,
emission from the disc's inner edge completely dominates the SED in
the mid--infrared, i.e. for $\lambda \sim 1$--4 to 10~$\mu$m.  For large
values of the accretion rate $\dot{M}_f$  exceeding about
$10^{-8}$~$M_{\sun}$~yr$^{-1}$, the emission predominantly comes from
the shocks   derived from fluid elements being propelled back towards  the disc.  For
lower accretion rates, irradiation from the central star becomes the
main source of emission from the inner edge.

When shocks dominate, we expect the SED to display some variability,
as the shock dissipation rate varies with time (see
fig.~[\ref{shockD}]) on a timescale of a few $\omega^{-1}$, where
$\omega$ is the angular velocity of the binary, i.e. comparable to the
binary period.  Even when irradiation dominates, we would expect some
variability on a timescale comparable to the binary period as the
distance between the edge of the disc and the stars varies as the
stars move  around their centre of mass.

\subsubsection* {Application to CoRoT~223992193:}

CoRoT~223992193 is a double-lined PMS eclipsing binary which was
discovered by Gillen et al. (2014).  Its age has been estimated to be
between 3.5 and 6 Myr.  The SED of this system presents a
mid--infrared excess between 4 and 10~$\mu$m.  Gillen et al. (2014)
found that this excess could not be produced by a circumbinary disc
unless it had an accretion rate of $10^{-7}$~$M_{\sun}$~yr$^{-1}$,
four orders of magnitude higher than the accretion rate derived from
the H$\alpha$ emission from the stars. It was then  proposed that the
excess may be due to thermal emission from dust present in the
vicinity of the two stars.  As little as $10^{-13}$~M$_{\sun}$ of dust
would be needed to produce the observed flux.  This dust may also be
responsible for the out--of eclipse variability observed in this
system (Terquem, S\o rensen--Clark \& Bouvier 2015).  

\noindent Here, we comment that the excess emission observed for
CoRoT~223992193 is very similar to what would be expected from the
inner edge of a circumbinary disc.  In this system, emission would not
be dominated by the shocks as the accretion rate onto the stars is
only $10^{-11}$~$M_{\sun}$~yr$^{-1}$ and the fiducial accretion rate
through the disc is probably not that much larger (as $H/r$ is not
expected to be smaller than 0.05 or so in a disc around a PMS binary).
Emission would then be produced by irradiation from the stars.  With
$H_{\rm in}=0.05$, the disc's inner edge would produce a flux with a
maximum of about $3.5 \times 10^{-13}$~erg~cm$^{-2}$~s$^{-1}$ at a
wavelength between 3 and 4~$\mu$m, and a flux about 3.5 times smaller
at 10~$\mu$m (see the lower plots on fig.~[\ref{fig3}]).  The maximum
value of the flux would be increased up to
$10^{-12}$~erg~cm$^{-2}$~s$^{-1}$ if we adopted $H_{\rm in}=0.15$
instead.  Such an enhanced value of the disc's aspect ratio near the
edge of the cavity is consistent with 3D MHD simulations (Shi \&
Krolik~2015) and may be produced by magnetic fields that get
relatively stronger in the cavity.  The peak of the excess measured
for CoRoT~223992193 is about
$3 \times 10^{-12}$~erg~cm$^{-2}$~s$^{-1}$ at 4~$\mu$m and decreases
by the same factor as in our model at 10~$\mu$m (see fig.~[13] of
Gillen et al.~2014).  Given that we have calculated irradiation of the
disc's inner edge in a very crude way, our results are close enough to
the observations that emission from the disc's inner edge appears as a
serious candidate for producing the observed excess.  We note that, as
the system is seen with an inclination $i=85^{\circ}$ (Gillen et
al. 2014) which, although large, is such that $90^{\circ}-i$ in
radians exceeds $ H/r$, accordingly the inner rim of the disc would
not be obscured by the disc itself.

\subsubsection*{Application toV1481~Ori:}

V1481~Ori is a spectroscopic binary with a period of about 4.4 days in
the Orion nebula cluster.  It is associated with a strong infrared
excess which suggests the presence of an accreting circumbinary disc
(Messina et al. 2016 and references therein).  Recent spectroscopic
and photometric observations performed by Messina et al. (2016) in the
$V$--band at $\sim 0.7$~$\mu$m and in the $I$--band have shown that
the luminosity ratio of the stellar components varies in a way that
can be explained by the presence of a hotspot on the secondary, which
is speculated to be produced by accretion from a circumstellar disc.
These observations also show that the minimum value of the luminosity
ratio is 30\% larger than what would be expected in the case of no
accretion.  Messina et al. (2016) conclude that there has to be an
additional source of emission which is visible at all phases.

\noindent Here, we suggest that this excess of energy may be produced
by the inner edge of the circumbinary disc.  Given the short
wavelength at which this excess occurs, it would have to be produced
by shocks localised over a small fraction of the surface at the inner
edge, as illustrated on the upper right plot of figure~(\ref{fig3}).
The peak of the excess on this plot is at about 1~$\mu$m, so that
there is significant emission in the bands in which Messina et
al. (2016) have made the observations.  This model requires a rather
large accretion rate in the circumbinary disc which, as pointed out
above, is suggested by the observations.

\section*{Acknowledgements}

It is a pleasure to thank Julian Krolik for stimulating discussions
about accretion onto binary systems.

%
%===========================================================
%

%\bibliographystyle{apj}
%\bibliographystyle{plain}
%\bibliographystyle{mn2e}
%\bibliography{biblio_papers}

%
%===========================================================

\label{lastpage}
\end{document}